\documentclass[preprint,notitlepage,a4paper,showpacs,showkeys]{revtex4-1}

\usepackage{graphicx}
\usepackage[font=normalsize,labelsep=period]{caption}
\usepackage{dcolumn}
\usepackage{amsmath,amssymb,mathtools}
\usepackage{epstopdf}

\newcommand{\rf}[1]{(\ref{#1})}

\newcommand{\beq}[1]{\begin{equation} \label{#1}}
\newcommand{\eeq}{\end{equation}}
\newcommand{\beqa}[1]{\begin{eqnarray} \label{#1}}
\newcommand{\eeqa}{\end{eqnarray}}
\newcommand{\beqn}{\begin{eqnarray*}}
\newcommand{\eeqn}{\end{eqnarray*}}

\newcommand{\veps}{\varepsilon}

\newcommand{\ga}{\alpha}
\newcommand{\gb}{\beta}
\newcommand{\gl}{\lambda}

\newcommand{\go}{\omega}

\newcommand{\gD}{\Delta}

\renewcommand{\bm}{{\bf m}}

\newcommand{\const}{{\rm const}}

\begin{document}

\title{Cosmological dynamics based on Lovelock's gravity. Qualitative analysis}

\author{A.~V. Nikolaev}\email[]{ilc@xhns.org}
\affiliation{Laboratory of Gravity, Cosmology, Astrophysics, I.N.~Ulyanov Ulyanovsk State Pedagogical University, 4/5 Lenin Square, Ulyanovsk, 432071, Russia}
\author{V.~M. Zhuravlev}
\affiliation{Laboratory of Gravity, Cosmology, Astrophysics, I.N.~Ulyanov Ulyanovsk State Pedagogical University, 4/5 Lenin Square, Ulyanovsk, 432071, Russia}
\author{S.~V. Chervon}\email[]{chervon.sergey@gmail.com}
\affiliation{Laboratory of Gravity, Cosmology, Astrophysics, I.N.~Ulyanov Ulyanovsk State Pedagogical University, 4/5 Lenin Square, Ulyanovsk, 432071, Russia}
\affiliation{Department of Physics, Bauman Moscow State Technical University, 2nd Baumanskaya Street, 5, Moscow, 105005, Russia}

\date{\today}

\begin{abstract}
We perform a complete qualitative analysis of the cosmological dynamics of Lovelock gravity in a spatially flat Friedmann--Robertson--Walker (FRW) universe filled with a perfect fluid obeying a barotropic equation of state $p=\omega\rho$. Starting from the generalized $N$-dimensional Friedmann equations written in terms of the independent components of the Riemann tensor, we reduce the dynamics to a single autonomous first-order equation for the Hubble parameter $H$, whose right-hand side is a ratio of polynomials in $H^2$ fixed by the order $n$ of the Lovelock polynomial, the number of dimensions $N$, and the coupling constants $\alpha_i$. The real roots of these polynomials determine the fixed points and the singular points of the system, which govern its asymptotic behaviour. We give a complete classification of the possible evolution scenarios, identify the attractive and repulsive fixed points, locate the ``phantom intervals'' in which the effective matter density becomes negative, and show that fixed points are reached in infinite time whereas singular points are reached in finite time. For $N>4$ and suitable negative couplings a qualitatively new scenario arises, which we call the ``Big Shock'': the universe starts from a state with finite density and finite scale factor but with an infinite rate of change of the Hubble parameter, replacing the standard Big Bang. The general analysis is illustrated for $n=1,\,N=4$ (general relativity), $n=2,\,N=5$ (Einstein--Gauss--Bonnet gravity), and $n=3,\,N=7$ (cubic Lovelock gravity).
\end{abstract}


\maketitle

\section{Introduction}
Generalized (modified) theories of gravity, such as scalar-tensor, $f(R)$-gravity, Einstein-Gauss-Bonnet~(EGB) and Lovelock gravity, play a key role in modern cosmological studies, offering a natural generalization of general relativity~(GR)~\cite{1,2,3}; for comprehensive reviews of modified gravity, including the EGB and Lovelock-type models and the unified description of the early- and late-time cosmic history they provide, see \cite{NO2011,NOO2017}.
Such theories include corrections that take into account additional gravitational scalar fields, the consequences of quantum effects, and additional dimensions, which are especially important for describing the early universe, dark energy, and anomalous gravitational phenomena.
For example, the Einstein-Gauss-Bonnet theory (EGB), a special case of Lovelock theory, makes it possible to explain the accelerated expansion of the Universe
without involving the $\Lambda$CDM model, as well as to investigate the structure of space-time in conditions of high curvature~\cite{4,5,6}.

In the work~\cite{7}, generalized Friedmann equations for the $N$-dimensional Lovelock theory were derived. For a flat universe ($k=0$), non-trivial vacuum solutions of the (anti-)de Sitter type were found, depending on the parameters of the theory: the order of the Lovelock polynomial $n$, the number of dimensions $N$, and the coupling constants $\alpha_i$. These solutions demonstrate that the Lovelock theory allows for expanding solutions even in the absence of matter, which opens up the possibility of alternative modeling of cosmological dynamics. To derive the general field equations of Lovelock's gravity theory, we use an approach based on the independent components of the Riemann tensor, which has proven effective in studying the geometric properties of spacetime and finding exact solutions~\cite{8,11}. In~\cite{9,10}, it is shown how the static condition and spherical symmetry allow us to find solutions for charged dust in the EGB theory.

However, the behavior of solutions in the presence of matter, especially in the case of a perfect fluid, remains understudied. A dynamical analysis of a spatially flat universe with an equation of state $p=\omega \rho$ allows one to:
\begin{itemize}
	\item investigate the stability of Lovelock's theory solutions in comparison with general relativity;
	\item determine the role of the additional Lovelock terms in the evolution of the scale factor $a(t)$;
	\item understand how the theory parameters ($n$,$N$,$\alpha_i$) affect cosmological dynamics.
\end{itemize}

To achieve these goals, we use the methods of qualitative analysis of dynamical systems.

In this work, we focus on the cosmological implications of Lovelock's theory of gravity (LTG). In recent years, the issues of compactification, singularities, and anisotropy in multidimensional models have been actively explored within the framework of LTG. A vast body of work has been devoted to the general aspects of LTG, including exact and radiative solutions, black holes, thermodynamics, generalizations, and $f(\mathrm{Lovelock})$-models, see, for example, \cite{Dehghani2009,2,Deser2012,Dadhich2013,Padmanabhan2013,Chakraborty2015,Kong2025,Bueno2016,Concha2017}. Lovelock-type structures also arise in braneworld cosmology, where geodetic brane models supplemented by Gibbons--Hawking--York--Myers boundary terms lead to second-order Friedmann-type equations \cite{Arroyo2026}.

A summary of the current state of multidimensional cosmology in LTG is given in the review \cite{Pavluchenko2024}, where it is shown for a wide range of models (EGB, cubic Lovelock; vacuum, $\Lambda$-term, perfect fluid, spatial curvature and their combinations) that an initially anisotropic space naturally evolves into a configuration with two isotropic subspaces, and a realistic compactification of additional dimensions is possible at $D\ge 2$ for EGB with vacuum or $\Lambda$, always --- for vacuum cubic Lovelock, and at $D\ge 4$ --- for EGB with spatial curvature.

In \cite{Bousder2023}, solutions for early inflation and late accelerated expansion are obtained through the Friedmann equations within Lovelock quantum gravity; the renormalized Lovelock constants are interpreted as the topological mass of the corresponding branch, and the predictions for the spectral index and tensor-scalar ratio are expressed through the renormalized Hubble parameter and the number of $e$-folds.

Cosmological aspects of LTG in more particular settings are considered in recent works \cite{7,BrasselSubmitted,SinghBrasselMaharaj2025,BrasselSinghMaharaj2025}. In \cite{7}, universal cosmological solutions of Lovelock theory for the spatially flat Friedmann--Robertson--Walker (FRW) metric are constructed. In \cite{SinghBrasselMaharaj2025} and \cite{BrasselSinghMaharaj2025}, respectively, dust models and models with a dark energy equation of state in multidimensional Gauss-Bonnet gravity in maximally symmetric spacetime are studied; it is shown that the dynamics are governed by the second-kind Abel equation, and explicit solutions for the scale factor are obtained for arbitrary dimension and spatial curvature. In \cite{BrasselSubmitted}, the equation of state of vacuum dark energy for a FRW universe in Lovelock gravity with effective thermodynamics is analyzed. Beyond cosmology, exact interior solutions for static stars in third-order Lovelock gravity have been obtained recently \cite{NaickerMaharajBrassel2026,MaharajNaickerBrassel2026,NaickerBrasselMaharaj2026}; in particular, the stellar models of \cite{NaickerMaharajBrassel2026} are constructed in the seven-dimensional third-order theory, whose cosmological dynamics is analyzed in detail in the present paper.

The study of cosmological models using the methods of qualitative analysis of dynamical systems has been used for quite a long time, as reflected in the monograph by Bogoyavlensky \cite{Bogoyavlensky1980}. The active use of qualitative analysis of dynamical systems in models of cosmological inflation began with the pioneering work of A.A. Starobinsky \cite{Starobinsky} and the work of V.A. Belinsky and his colleagues \cite{Belinsky1985}.

Currently, this approach is used to analyze cosmological dynamics in models with various types of material sources of gravity, see, for example, the works \cite{Copeland2006}, \cite{BCNO2012}, \cite{ZhPP11GC}, \cite{Ign1,Ign2,Zh16STFI}, \cite{Duchaniya2024}, and the literature cited therein. Phenomenological modified-gravity descriptions of the late-time acceleration are also actively developed in other geometric frameworks, e.g.\ in $f(Q)$ gravity \cite{MyrzakulovKoussourGogoi2023}. These works have extensively studied the asymptotic properties of cosmological models with both perfect fluid and scalar field in various formulations. The paper \cite{zhch-2020} presents a qualitative analysis of the dynamics of a chiral cosmological model (CCM) with two scalar fields in a spatially flat FRW universe. The methods of qualitative analysis for cosmological models with degenerate fermions, Higgs scalar fields, and related processes were discussed in a series of papers by Yu. Ignatiev and his colleagues \cite{iug-2}-\cite{iug-5}. Exactly solvable and integrable FRW models constitute a complementary line of this research \cite{Esmakhanova2011}. Phase-plane and dynamical-systems techniques of the same kind are also applied beyond cosmology, e.g.\ to the temporal evolution of radiating stars in general relativity and in Einstein--Gauss--Bonnet gravity \cite{MaharajGovinder2025,BogadiEtAl2026,MaharajBrasselSingh2026}.

An important aspect of these works is the ability to establish all the main asymptotics of the models without having exact solutions. The basis of the qualitative analysis is a simple strategy for calculating and classifying all the singular points of the dynamical systems representing the model under consideration. The asymptotic properties of the model parameters are usually analyzed using perturbation theory for linearized equations \cite{Bogoyavlensky1980,BL89}, which allows for analytical research.  Accompanying this qualitative information with phase diagrams constructed using numerical solutions of the equations of cosmological models, it is possible to obtain a very visual way of analyzing the evolution processes of the universe within the framework of the assumptions made.
In addition, in some situations, by extending the standard methods of qualitative analysis \cite{Bogoyavlensky1980,BL89}, it is possible to obtain additional information about the nature of the universe's evolution. As an example, we can cite the work \cite{zhch-2020}, where the elements of the phase portraits of the system are special curves and surfaces associated with the properties of the internal metric of the chiral space. In the case of the Lovelock models considered in this work, additional elements of the phase diagrams are singular points, which are associated with a new type of possible initial state of the universe.

The main objective of this work is to perform a complete qualitative analysis of Lovelock cosmological models with a flat FRW metric and a spacetime dimension of $4\le D \le 7$. The work provides a complete classification of all possible scenarios for the evolution of the universe within these models for almost arbitrary values of the coupling coefficients of Lovelock's theory of gravity.

Section 2 presents the general equations of Lovelock gravity theory and their form in terms of independent components of the Riemann tensor in the FRW metric. A perfect fluid is considered as the source of gravity. Section 3 presents the equations of cosmological dynamics in a spatially flat universe for a barotropic fluid, which reduce to a single equation in autonomous form. The resulting equation for the Hubble parameter represents an autonomous dynamical system for any parameters of the Lovelock polynomial $n$, the number of dimensions 
$N$, and expansion coefficients $\alpha_i$ $(i=2,...,n)$.

Section 4 provides a detailed qualitative analysis of the aforementioned autonomous dynamical system. Specifically, the relationship between the scale factor $a(t)$
and the Hubble parameter $H(t)$
is established, an analysis of small deviations from the fixed points of the system is presented, the attractive and repulsive fixed points are identified, and regions with negative matter density are determined and discussed from the point of view of the energy conditions. Furthermore, the behavior of the system near singular points is studied, and the time to reach the fixed and singular points is found. Section 5 presents examples of the analysis of cosmological dynamics for Lovelock gravity for 
$n=1, N=4$ (which is equivalent to general relativity); for 
$n=2, N=5$ (Einstein–Gauss–Bonnet gravity); and for 
$n=3, N=7$ (Cubic Lovelock). In all cases, possible scenarios of the universe's evolution are considered. Section 6 summarizes the results of the study. Appendix~A provides the derivation of the relationship between the scale factor $a(t)$
and the Hubble parameter $H(t)$, while Appendix~B collects several exact statements about the polynomials $P_1$ and $P_2$ and their roots.

\section{Lovelock's theory of gravity and cosmology}

Lovelock gravity is a natural generalization of general relativity, and it is the most general theory of gravity that has second-order field equations:
\begin{equation}
    \mathcal{\prescript{(1)}{}H}_{ab} + \sum_{i=2}^{n}\alpha_i \mathcal{\prescript{(i)}{}H}_{ab} = T_{ab},
    \label{Lavelockfield}
\end{equation}
where the Latin indices run from $0, 1, \ldots, N-1$, $N$ is the number of dimensions, $n$ is the order of the Lovelock polynomial, $\alpha_i$ are arbitrary constants, $T_{ab}$ is the energy-momentum tensor, and $\mathcal{H}_{ab}$ is the Lovelock tensor:
\begin{equation}
    \mathcal{\prescript{(n)}{}H}_{ab} = n \mathcal{R}_{ab} - \frac{1}{2}\mathcal{R}g_{ab},
    \label{LHDef}
\end{equation}
where $\mathcal{R}_{ab}$ is the generalized Ricci tensor:
\begin{subequations}
    \begin{equation}
        \mathcal{R}_{ab} = g^{cd}\mathcal{R}_{cadb},
        \label{LRicciDef}
    \end{equation}
    \begin{equation}
        \mathcal{R} = g^{ab}\mathcal{R}_{ab},
        \label{LRicciSDef}
    \end{equation}
\end{subequations}
$\mathcal{R}_{abcd}$ is the generalized Riemann tensor:
\begin{equation}
    \mathcal{R}_{abcd} = Q_{ab}^{\ \ mn}R_{mncd}.
    \label{PureRdef}
\end{equation}
The tensor $Q^{ab}_{\ \ cd}$ is defined as follows:
\begin{equation}
    Q^{ab}_{\ \ cd} = \frac{1}{2^n}\delta^{aba_2b_2\dots a_{n} b_{n}}_{cdc_2d_2\dots c_{n} d_{n}} R_{a_2b_2}^{\ \ c_2d_2}\dots R_{a_{n}b_{n}}^{\ \ c_{n}d_{n}},
    \label{PureQdef}
\end{equation}
where $\delta^{aba_2b_2\dots a_{n} b_{n}}_{cdc_2d_2\dots c_{n} d_{n}}$ is the generalized Kronecker delta.

Consider the case of Friedmann space:
\begin{equation}
    ds^2 = -dt^2 + a(t)^2\left(\frac{1}{1 - \kappa r^2}dr^2 + r^2d\Omega_{N-2}^2\right),
    \label{metricN}
\end{equation}
where $r$ is the radial coordinate, and $d\Omega_{N-2}$ stands for the angular part of the metric. It was previously shown that the Riemann tensor has only two independent components~\cite{7}:
\begin{subequations}
    \begin{align}
        \label{notation5_f}
        f_1 =& R^{1\alpha}_{\ \ 1\alpha} = R^{\alpha\beta}_{\ \ \alpha\beta} = \frac{\dot{a}^2}{a^2} + \frac{\kappa}{a^2},\\
        f_2 =& R^{10}_{\ \ 10} = R^{\alpha 0}_{\ \ \alpha 0} = \frac{\ddot{a}}{a},
        \label{notation5_l}
    \end{align}
        \label{notation5}
\end{subequations}
where the Greek indices run from $2, 3, \ldots, N-2$. This allows us to express the field equations \eqref{Lavelockfield} in terms of the independent components of the Riemann tensor \eqref{notation5} using the energy-momentum tensor of a perfect fluid:
\begin{subequations}
    \begin{equation}
        -\left( N - 2 \right)\left( \frac{N - 3}{2}f_1 + f_2 \right) + \sum_{i=2}^n \alpha_i f_1^{i-1}\left( \prescript{(i)}{}k_{11}f_1 + \prescript{(i)}{}k_{12}f_2 \right) = p,
        \label{RFriedman1}
    \end{equation}
    \begin{equation}
        -\frac{1}{2}\left( N - 1 \right)\left( N - 2 \right)f_1 + \sum_{i=2}^n \alpha_i \prescript{(i)}{}k_{21}f_1^{i} = -\rho,
        \label{RFriedman2}
    \end{equation}
\end{subequations}
where $\prescript{(n)}{}k_{ij}$ are the combinatorial coefficients:
\begin{subequations}
    \begin{align}
        \label{eqK11}
	    \prescript{(n)}{}k_{11} &= - \frac{1}{2}\left(2\left( n - 1 \right)\right)!\left( N - 2 \right)\left( N - 1 -2n \right)C^{2\left( n - 1 \right)}_{N - 3},\\
	    \prescript{(n)}{}k_{12} &= -\frac{1}{2}\left(2\left( n - 1 \right)\right)!\left[ 2\left( N - n - 1 \right)C^{2\left( n - 1 \right)}_{N - 2} + \left( N - 2 \right)\left( N - 1 -2n \right)C^{2n - 3}_{N - 3} \right],\\
	    \prescript{(n)}{}k_{21} &= - \frac{1}{2}\left(2\left( n - 1 \right)\right)!\left( N - 1 \right)\left( N - 2 \right)C^{2\left( n - 1 \right)}_{N - 3},\\
	    \prescript{(n)}{}k_{22} &= \frac{1}{2}\left(2\left( n - 1 \right)\right)!\left( N - 1 \right)\left[ 2\left( n - 1 \right)C^{2\left( n -1 \right)}_{N - 2} - \left( N - 2 \right)C^{2n - 3}_{N - 3} \right] = 0.
        \label{eqK22}
    \end{align}
        \label{eqK}
\end{subequations}
The symbol $!$ denotes the factorial, and $C^n_m = m!/\left(n!\,(m-n)!\right)$ is the binomial coefficient (the number of combinations of $m$ elements taken $n$ at a time). Note that the combinatorial coefficients in \eqref{eqK} depend only on $n$ and $N$ and are numbers. For each order $i$ entering the sums in \eqref{RFriedman1}--\eqref{RFriedman2}, the coefficients $\prescript{(i)}{}k_{lm}$ are obtained from \eqref{eqK} by the substitution $n\to i$; note that they depend on the summation index, so the coefficients of different Lovelock orders enter the equations with their own combinatorial weights.

\section{Equations of cosmological dynamics}

In the case of a spatially flat universe, $\kappa = 0$, the functions $f_1$ and $f_2$ are as follows:
$$
    f_1 = H^2.~~~f_2 = \dot{H} + H^2,
$$
where $H=\dot{a}/a$ is the Hubble parameter.
Using the algebraic relation for \eqref{eqK}:
$$
k_{11} + k_{12} - k_{21} = 0,
$$
the dynamics equations are reduced to:
\begin{subequations}
	\begin{equation}
		P_1(H^2)\dot{H} + H^2P_2(H^2) = p,
		\label{flatF1}
	\end{equation}
	\begin{equation}
		H^2P_2(H^2) = -\rho,
		\label{flatF2}
	\end{equation}
	\label{flatF}
\end{subequations}
where the following definitions of polynomials were introduced:
\begin{subequations}
	\begin{equation}
		P_1 =-\Big( N-2 \Big) + P_{10}(H^2),
		\label{P1}
	\end{equation}
	\begin{equation}
		P_2 = -\Big( N - 1 \Big)\Big( N - 2 \Big)/2+P_{20}(H^2),
		\label{P2}
	\end{equation}
    \begin{equation}
		P_{10} = \sum_{i=2}^{n}\alpha_i\, {}^{(i)}k_{12}\, H^{2i-2},\qquad
        P_{20} = \sum_{i=2}^{n}\alpha_i\, {}^{(i)}k_{21}\, H^{2i-2}.
		\label{P0}
	\end{equation}
\end{subequations}

In the case of a barotropic equation of state
\begin{equation}
    p=\omega(t) \rho,
	\label{Eqprho}
\end{equation}
the system of equations \eqref{flatF1} and \eqref{flatF2} is transformed into a single equation of the following form:
\begin{equation}
	\dot{H} = - \Gamma(t) H^2\frac{P_2(H^2)}{P_1(H^2)}=F(H).
	\label{dotH}
\end{equation}
Here:
$$
	\Gamma = 1+\omega(t).
$$
The dependence of the barotropic parameter $\omega(t)$ on time, in particular, describes situations when the Universe is filled with matter consisting of a mixture of various components \cite{Zh01JETP}, which is especially important in the early Universe. This model can also be related to models with a scalar field. If the function $1+\omega(t)$ is \textbf{sign-constant}, then equation \eqref{dotH} can be reduced to an autonomous equation:
\begin{equation}
	\frac{dH}{d\tau} = - \varepsilon H^2\frac{P_2(H^2)}{P_1(H^2)}=F(H),
	\label{dotHt}
\end{equation}
where $\varepsilon={\rm sign}\Big(\Gamma(t)\Big)$ and a new variable $\tau$ is introduced:
$$
    \tau = \int\limits_{0}^{t} \Big|\Gamma(t')\Big| dt'
$$
This allows us to use qualitative analysis methods to study the general properties of this equation \cite{BL89}.
In the case where the function $\Gamma(t)$ changes sign during the evolution of the universe and has zeros within the time of evolution, the entire time interval should be divided into subintervals with a fixed sign of $\Gamma(t)$. In each subinterval, the equation will be autonomous. Let us note what can be said about the sign-changing case in general. If $\omega$ is a given function of the scale factor, $\omega=\omega(a)$ (a mixture of components with different dilution laws \cite{Zh01JETP}, or a phenomenological parametrization crossing the phantom divide $\omega=-1$ \cite{Vikman2005,CaiQuintom2010}), one may introduce instead of $\tau$ the variable $d\sigma=\Gamma(a)\,d\ln a$, in terms of which \eqref{dotH} takes the form $dH/d\sigma=-H\,P_2(H^2)/P_1(H^2)$ and does not contain $\Gamma$ at all. Hence the ``skeleton'' of the phase line --- the fixed points $\pm h_i$, the singular points $\pm s_i$, the basins between them and the intervals of negative density introduced below --- is determined by the geometry ($n$, $N$, $\alpha_i$) alone and does not depend on the equation of state. A change of the sign of $\Gamma$ reverses the direction of motion along the same phase curve for all $H$ simultaneously (attracting fixed points become repelling and vice versa): at the moment when $\omega$ crosses $-1$ (away from the singular points, where $P_1=0$) the Hubble parameter passes through an extremum and the representative point retraces its path along the phase line (the same values of $H$ are passed in the reverse order, at later times and larger $a$), and it never leaves the basin in which it started, since the boundaries of the basins are either fixed points, reached only asymptotically, or singular points, at which the evolution terminates. In particular, in the models with singular points a sufficiently long phantom phase ($\Gamma<0$) inside the central basin drives the trajectory towards the singular point $s_1$ rather than to infinity, so that the Big Rip of general relativity is replaced by the Big Shock discussed below. If, on the other hand, the equation of state is given in the form $p=p(\rho)$, then, since $\rho=\rho(H)$ by \eqref{flatF2}, one has $\Gamma=\Gamma(H)$: equation \eqref{dotH} remains autonomous, but the zeros of $\Gamma(H)$ become additional fixed points of the system, and their location with respect to the points $h_i$ and $s_i$ is determined entirely by the specific form of $\Gamma(H)$. Formally the two descriptions are equivalent, but they correspond to different physical situations, and each choice of $\omega(a)$ or $\Gamma(H)$ produces its own phase diagram; a detailed analysis of these cases is beyond the scope of this article. Hereinafter, we will consider the variant of the sign-constant function $\Gamma(t)$, which, after the transformation $t\to\tau$, is practically the same as the case $\omega={\rm const}$.

Under the above constraints, the equation \eqref{dotHt} represents a one-dimensional dynamical system with respect to the Hubble parameter $H$ for any admissible Lovelock geometry parameters $n$, $N$ and the coupling coefficients $\alpha_i,~i=2,\ldots,n$. The right-hand side of \eqref{dotH} is in general a ratio of polynomials depending on $H^2$. This allows us to provide a general understanding of the possible types of cosmological evolution in the spatially flat Lovelock universe based on a qualitative analysis of the dynamical system \eqref{dotH}.

We note that the restriction to the spatially flat case is made for simplicity. For $\kappa\neq0$ the field equations \eqref{RFriedman1}--\eqref{RFriedman2} are expressed through the same polynomials $P_1$ and $P_2$ of the argument $H^2+\kappa/a^2$ (cf.\ the braneworld analyses, where this combination likewise enters as a single argument, $\chi=\sqrt{H^2+\kappa/a^2}$ \cite{Arroyo2026}), so that in the physical region $H^2+\kappa/a^2\ge0$ the roots $s_i$ and $h_i$ retain their meaning for the variable $\chi$; the dynamical system, however, becomes two-dimensional in the variables $(a,\chi)$, and its qualitative analysis will be presented elsewhere.

\section{Qualitative analysis of the dynamical system}

To investigate the asymptotic properties of the dynamical system \eqref{dotH}, we represent the polynomials on the right-hand side of this equation as follows:
\begin{equation}
	 P_1 = A_1\prod\limits_{i=1}^{n-1}(H^2-s_i^2),~~
     P_2 = A_2\prod\limits_{i=1}^{n-1}(H^2-h_i^2),~
	\label{DefPP}
\end{equation}
Here, $s_i^2$ are the roots of the polynomial $P_1(H^2)$, and $h_i^2$ are the roots of the polynomial $P_2(H^2)$, and the constants $A_1=\alpha_n\,{}^{(n)}k_{12}$ and $A_2=\alpha_n\,{}^{(n)}k_{21}$ are the leading coefficients of the polynomials $P_1(Z)$ and $P_2(Z)$, $Z=H^2$.
According to the general theory of qualitative analysis of dynamical systems \cite{BL89}, the general asymptotic dynamics of the model is determined by the type and location of the roots $h_i^2$ and $s_i^2$. Only the roots that satisfy the conditions $h_i^2>0$ and $s_i^2>0$ are of interest. The roots $h_i^2>0$ of the polynomial $P_2(H^2)$ in the numerator of the right-hand side of \eqref{dotH} determine the position of the fixed points of the system \eqref{dotH}. The roots $s_i^2$ of the polynomial $P_1(H^2)$ in the denominator of the right-hand side of \eqref{dotH} determine the points of unlimited growth. These points will be referred to as singular points. The negative and complex roots $h_i^2$ and $s_i^2$ are not associated with the singular points of the dynamical system and do not affect its asymptotic behavior. Since the right-hand side of \eqref{dotH} depends on $H^2$, the singular points of the system are pairs of values of the Hubble parameter $H$ that differ in sign: $s_i = \pm |s_i|$ and $h_i=\pm |h_i|$, and are therefore symmetrically located relative to the point $H=0$. The point $H=0$ itself is an equilibrium point of a special, degenerate kind (its status is analyzed in a separate subsection below); it should not be confused with the singular points $s_i$, at which the right-hand side of \eqref{dotH} diverges. We will denote it by $h_0$.

\subsection{Formal dependence of the scale factor on the Hubble parameter} 

An interesting general property of Lovelock models is that, in the case of $\omega={\rm const}$, for an arbitrary model, the scale factor depends on the Hubble parameter. To show this, we require $\go=\const$ and rewrite equation \rf{dotH} as follows:
\begin{equation}
	\frac{d \ln a}{dH} = - \frac{1}{(1+\omega) H}\frac{P_1(H^2)}{P_2(H^2)}.
	\label{dotadH}
\end{equation}

Passing to the variable $Z=H^2$, we rewrite \eqref{dotadH} as
\begin{equation}
     \frac{d \ln a}{dZ} = - \frac{M}{Z}\frac{P_1(Z)}{P_2(Z)},\qquad M=\frac{1}{2(1+\go)}.
\label{dotadZ}
\end{equation}
The general solution of \eqref{dotadZ} can be written in closed form (see Appendix~A):
\begin{equation}
	\ln(a/a_0) = -\frac{M}{A_2}\Big(J_0(H^2)+Q(H^2)+\sum\limits_{i=1}^{n-1}K_iP_1(Z_i)\ln|H^2-Z_i|\Big),\qquad Z_i=h_i^2,
	\label{Sola}
\end{equation}
where
\beqn
     &&J_0(Z) = K_0\Big(\gb_0\ln|Z|+\sum\limits_{j=1}^{n-1}\gb_j\frac{Z^j}{j}\Big),\qquad
     Q(Z) = \sum\limits_{i=1}^{n-1}K_i\sum\limits_{j=1}^{n-1}\gb_jT_{j}(Z,Z_i),\\
     &&T_j(Z,Z_i)=\sum\limits_{k=1}^{j}\frac{Z^k}{k}Z_i^{j-k},\qquad
     \gb_0=-(N-2),\qquad \gb_j=\alpha_{j+1}\,{}^{(j+1)}k_{12},~~j=1,\ldots,n-1,
\eeqn
Here $\gb_j$ are the coefficients of the polynomial $P_1(Z)=\sum_{j=0}^{n-1}\gb_jZ^j$, $A_2$ is the leading coefficient of $P_2$, and the constants $K_0,\ldots,K_{n-1}$ are determined by the partial-fraction decomposition described in Appendix~A.

It should be noted that the relation \rf{Sola} is obtained under the assumption that the polynomial $P_2(Z)$ has no multiple roots. A general criterion for the multiplicity of the roots of $P_2(Z)$ for arbitrary $N$ and $n$ will not be discussed here because of the complexity of such computations; it is simpler to state the corresponding criteria for each particular choice of $N$ and $n$, as is done below. Let us only note that the multiplicity of the roots affects solely the procedure of computing the function $a=a(H)$ and does not touch upon the other aspects of this work. Note also that a coincidence of some of the roots of the polynomials $P_1(Z)$ and $P_2(Z)$ reduces the behaviour of the system to that of models with a smaller value of $n$ than the current one, so that such a coincidence does not change the general classification of the possible evolution scenarios. Finally, the relation \rf{Sola} gives the dependence $a=a(H)$ rather than the explicit dynamics $a=a(\tau)$, and the growth of the degrees of the polynomials with $n$ makes its analysis increasingly cumbersome. For $\veps=-1$ the sign of the right-hand side of \eqref{dotadZ} is reversed, so the exponents in \eqref{Sola} change sign accordingly.

\subsection{Analysis of small deviations from fixed points in the system}

Let $h_i\not=0$ be some real root of the polynomial $P_2(H^2)$. Consider small deviations of the parameter $H$ from $h_i$, assuming:
$$
H = h_i + \xi(\tau),
$$
where $|\xi|<<1$. Since the right-hand side of equation \eqref{dotH} depends on $Z=H^2$, its Taylor expansion in the vicinity of the point $H=h_i$ is evaluated at the point $z_i=h_i^2$. In this case, taking into account that, up to second-order terms, $z =H^2\simeq h_i^2 + 2h_i\xi(\tau)$, we find:
$$
     F(z) = F(z_i) + 2h_i\left.\frac{dF(z)}{dz}\right|_{z=h^2_i}\xi(\tau)+O(\xi^2).
$$
The first derivative of the function $F(z)$ has the following form:
$$
     \frac{dF(z)}{dz} = - \varepsilon\frac{d}{dz}\left( z\frac{P_2(z)}{P_1(z)}\right)=
     - \varepsilon \left(\frac{P_2(z)}{P_1(z)}+z\frac{P'_2(z)}{P_1(z)}-z\frac{P_2(z)P'_1(z)}{P^2_1(z)}\right).
$$
Taking into account that $P_2(z_i)=0$, we obtain the following expression:
$$
    \left.\frac{dF(z)}{dz}\right|_{z=h^2_i} = - \varepsilon h_i^2\frac{P'_2(h_i^2)}{P_1(h_i^2)}.
$$
Then equation \eqref{dotH}, to first order in perturbation theory, can be transformed to the following general form:
\begin{equation}
	 \frac{d\xi}{d\tau} = 2h_i\left.\frac{dF(z)}{dz}\right|_{z=h^2_i}\xi = \gl_i \xi,
	\label{Eqh}
\end{equation}
where
$$
    \gl_i = -2\varepsilon h_i\sigma(h_i),~~~
    \sigma(h_i) = \frac{h^2_iP'_2(h^2_i)}{P_1(h^2_i)}.
$$
The equation \rf{Eqh} has a general solution of the following form:
\begin{equation}
	 \xi = A e^{\gl_i \tau}.
	\label{Solh}
\end{equation}
Since this solution is obtained under the assumption $|\xi|<<1$, then $|A|<<1$.
A fixed point is called stable or attractive if $\gl_i < 0$, and unstable (repulsive) if $\gl_i > 0$. In the case of $\gl_i = 0$, the stability analysis should be performed in the second order of perturbation theory. This special case will not be considered in this paper. Near a stable ($\gl_i<0$) fixed point $h_i$ with small initial deviations set by the constant $A: ~|A|<<1$, the system returns to the point $h_i$. In the case of an unstable fixed point ($\gl_i>0$), the system moves away from $h_i$ for any value of $A$.
Since the parameter $\gl_i$ contains the multiplier $h_i$, if the special point $h_i$ is stable, then the point $-h_i$ will be unstable, and vice versa, if $h_i$ is unstable, then $-h_i$ is stable.

\subsection{Attracting fixed points}

Near the stable point $H\simeq h_i$, if $h_i > 0$, the cosmological evolution of the scale factor differs slightly from the de Sitter expansion regime. Indeed:
$$
     H = \frac{d\ln a}{d\tau } = h_i + \xi(\tau).
$$
From here, taking into account \eqref{Solh}, we find:
\begin{equation}
     a \simeq a_0 e^{h_i \tau} \left(1 - \frac{A}{|\gl_i|}e^{-|\gl_i| \tau}\right).
\label{Solazp}
\end{equation}
Here, it is assumed that $|A|<<1$.
This relation indicates that near a fixed point with $h_i >0$, since $\lim\limits_{\tau\to\infty}e^{-|\gl_i| \tau} \to 0$, the solution for the scale factor grows exponentially, which is the definition of the de Sitter evolution regime as $\tau\to\infty$.

In the case of $h_i <0$, provided that this point is stable, i.e., $\gl_i < 0$, the corresponding solution for the scale factor describes an exponentially rapid contraction of the universe. In this case, the solution can be written as follows:
\begin{equation}
     a \simeq a_0 e^{-|h_i| \tau} \left(1 - \frac{A}{|\gl_i|}e^{-|\gl_i| \tau}\right).
\label{Solazm}
\end{equation}

It is also assumed that $|A|<<1$.

Let us pay attention to the physical effects of the behavior of the Universe near the stationary singular points $h_i$. According to \eqref{flatF2}, at the stationary points $H=h_i$ the density of matter
\begin{equation}
	\rho(H)=-H^2P_2(H^2)
	\label{DefrhoH}
\end{equation}
in the Universe goes to zero. In the case $h_i>0$, this fact reflects the obvious decrease in the matter density as a result of the de Sitter expansion. However, in the case $h_i<0$, the universe contracts as it approaches this point, as described by \eqref{Solazm}, but the matter density still tends to zero.

In fact, we come to the paradoxical behavior of the universe's matter near the attracting points $H=h_i<0$. This behavior is a property of Lovelock's cosmological models themselves. It manifests itself in models with parameters $\alpha_{i},~i=2,\ldots,n$ where the polynomial $P_2(z)$ has positive real roots. Since the argument of $P_2$ in the expression for the density and in the evolution equation is $H^2$, it is inevitable that any real root of the polynomial $P_2(z)$ will give two fixed points $\pm h_i$. In this case, the fixed repulsive point with $h_i>0$ always has a mirror-image attractive point of $-h_i$, which leads to a paradoxical decrease in the density as the universe contracts.

\subsection{Repulsive fixed points}

Near unstable points with $\gl_i>0$, the solutions for the scale factor in the first order of perturbation theory are as follows:
$$
     a \simeq a_0 e^{\pm |h_i| \tau} \left(1 + \frac{A}{\gl_i}e^{\gl_i \tau}\right).
$$
Based on this relationship, one would expect the asymptotic solution for the scale factor to be determined by the ratio of the growth rates of the exponentials with exponents $\pm |h_i|$ and $\gl_i >0$. However, as the small perturbations $\xi(\tau)$ grow exponentially, the condition of smallness of this quantity is no longer met quickly. As a result, the approximation no longer reflects the behavior of the exact solution of the equation. For a more detailed and visual analysis of the overall behavior of a system in such situations, it is more convenient and useful to use a graphical analysis of its phase trajectories, which will be presented in the following examples.

\subsection{The degenerate fixed point $H=0$}

The point $H=0$ requires a separate discussion, since the linear analysis of the previous subsections does not apply to it. Writing the right-hand side of \eqref{dotH} as $F(H)=-\veps H^2 P_2(H^2)/P_1(H^2)$, we find $F(0)=0$ and $F'(0)=0$, while
$$
     F''(0) = -2\veps\,\frac{P_2(0)}{P_1(0)} = -\veps\,(N-1) \neq 0 ,
$$
so that $H=0$ is a degenerate (nonhyperbolic) fixed point. In its neighborhood the dynamics is universal for all Lovelock models: since $P_2(0)/P_1(0) = (N-1)/2$ independently of the couplings $\alpha_i$, equation \eqref{dotH} reduces to
\beq{H0local}
	\frac{dH}{d\tau} \simeq -\frac{\veps\,(N-1)}{2}\, H^2 ,
\eeq
which coincides with the general-relativistic behavior: near $H=0$ the higher-order Lovelock terms are negligible and every model of the family reduces to \rf{HGR}. It follows from \eqref{H0local} that $H=0$ is semi-stable: for $\veps=+1$ it attracts the trajectories with $H>0$ and repels those with $H<0$, and vice versa for $\veps=-1$. The solution of \eqref{H0local}, $H \simeq 2/[\veps(N-1)(\tau-\tau_0)]$, shows that the trajectories approach $H=0$ only asymptotically, in an infinite time $\tau$, and never cross it: the expanding and the contracting branches of the evolution are dynamically disconnected. For this reason we refer to $H=0$ as a special (equilibrium) point and denote it by $h_0$, reserving the term ``singular points'' for the points $H=\pm s_i$, at which the right-hand side of \eqref{dotH} diverges.

\subsection{Areas with negative matter density}

The presence of real roots of the polynomial $P_2(H^2)$, with a suitable choice of the parameters $\alpha_{i},~i=2,\ldots,n$, leads to another feature of the Lovelock models for $n>2$ and $N>5$. Since at the points $h_i$ at which the polynomial vanishes, the sign of this polynomial changes under general conditions, at the same time the sign of the density of matter changes at these points as a function of the parameter $H$~\eqref{DefrhoH}.  Consequently, in Lovelock models, in the presence of real roots of the polynomial $P_2(H^2)$, there are regions of values of $H$ in which the density $\rho(H)$ becomes negative, which is physically impossible.
This means that in Lovelock models with matter in the form of a perfect fluid, there are intervals of values of $H$ in which these models are physically unrealizable. If a scalar field or a mixture of a scalar field and a perfect fluid is used instead of a perfect fluid in Lovelock models, then regions with negative energy density can be associated with phantom fields. Therefore, such intervals of $H$ values can be formally called phantom intervals. It should be noted that the boundary points of these intervals are fixed points, one of which is attractive and the other is repulsive. Moreover, a singular point $s_i$, which acts as a separatrix barrier, may lie inside such a phantom interval: at this point the matter density remains finite, while the acceleration diverges, $|\dot{H}|\to\infty$. As shown in Appendix~B, the central basin $(-s_1,s_1)$ is always free of phantom intervals, and if all $\alpha_i\ge 0$ phantom intervals are absent altogether.

\subsection{Energy conditions in the phantom intervals}
\label{SecEC}

The term ``phantom intervals'' requires a comment concerning the energy conditions. Since $\rho=-H^2P_2(H^2)$ and, by \eqref{flatF1}, $\rho+p=P_1(H^2)\dot H=\Gamma\rho$, the weak energy condition (WEC), $\rho\ge0$, $\rho+p\ge0$, is violated in these intervals by definition, whereas the null energy condition (NEC), $\rho+p\ge0$, depends on the sign of $\Gamma$: for $\omega>-1$ it is violated together with the WEC, while for $\omega<-1$ it is satisfied, since then $p=\omega\rho>|\rho|$. Thus the intervals are ``phantom'' in the sense of a negative energy density rather than in the sense of the phantom equation of state $\omega<-1$ at positive density \cite{C02}; the two notions have in common only that both require exotic matter. A more instructive form of the NEC follows from $\rho+p=P_1(H^2)\dot H$:
\begin{equation}
   \rho+p\ge0 \quad\Longleftrightarrow\quad P_1(H^2)\,\dot H\ge0 ,
   \label{NECgeom}
\end{equation}
i.e.\ away from the zeros of the two factors $\dot H$ and $P_1(H^2)$ must have the same sign. In the central basin $(-s_1,s_1)$, where $P_1<0$, this is the familiar condition $\dot H\le0$ of general relativity; beyond the first (simple) singular point, where $P_1>0$, the NEC requires $\dot H\ge0$ instead, and the sign is reversed again at every simple root of $P_1$. In other words, in the outer basins a super-accelerated expansion, $\dot H>0$, is compatible with matter satisfying the NEC: it is the effective kinetic coefficient of the theory that changes its sign, not the character of the matter.

The same situation can be described from the effective-fluid viewpoint, when all Lovelock corrections are transferred to the right-hand side of the Einstein-like equations (for the energy conditions in modified gravity in this formulation see \cite{CLM2014}). Then
\begin{equation}
   \rho_{\rm eff}=\frac{(N-1)(N-2)}{2}H^2\ge0,\qquad \rho_{\rm eff}+p_{\rm eff}=-(N-2)\dot H,\qquad
   \omega_{\rm eff}=-1+\frac{2\Gamma}{N-1}\,\frac{P_2(H^2)}{P_1(H^2)} ,
   \label{EffFluid}
\end{equation}
so that $\rho_{\rm eff}\ge0$ always, while the NEC, and hence the WEC, of the effective fluid hold if and only if $\dot H\le0$; furthermore, $\omega_{\rm eff}=-1$ exactly at the fixed points $h_i$, and $\omega_{\rm eff}\to\infty$ at the singular points $s_i$ (provided $\Gamma\rho\neq0$ there). For the effective fluid the Big Shock is therefore precisely a Type II (sudden) singularity, $p_{\rm eff}\to\infty$ at finite $a$ and $\rho_{\rm eff}$, in accordance with the discussion in Sec.~\ref{SecFTS} below. The phantom intervals are the regions where the Lovelock contribution $-H^2P_{20}(H^2)$ to \eqref{flatF2} alone exceeds $\rho_{\rm eff}$, so that the matter is forced to compensate it by a negative density. Our position is therefore twofold: for ordinary matter with $\rho\ge0$ the phantom intervals are excluded as dynamically unreachable (their boundaries are fixed points, see Sec.~\ref{SecGhosts}), whereas within the effective-fluid or phantom-scalar-field interpretation \cite{C02,GPRS06} they are admissible, the effective energy density being always non-negative and the energy conditions of the effective fluid being controlled by the sign of $\dot H$; the effective description, however, does not restore the WEC for the matter itself.

\subsection{The behavior of the system near singularity points}

Note that the right-hand side of equation \eqref{dotH} can be represented as follows:
\begin{equation}
	 F(H)=- \varepsilon H^2P_2(H^2)\sum\limits_{i=1}^{n-1}\left(\frac{B^{-}_i}{H-s_i}+\frac{B^{+}_i}{H+s_i}\right),
	\label{FP1}
\end{equation}
where the numerical coefficients $B_i,~B^{+}_i,~B^{-}_i$ are found by solving a system of linear algebraic equations in analogy with \eqref{P1} (see Appendix~A).
Thus, at the points $H=\pm s_i$ with $s^2_i>0$, the right-hand side $F(H)$ of equation \eqref{dotH} has poles. The poles can be attractive or repulsive.
When approaching these poles at the points $H=s_i$, $|\dot{H}|
\to \infty$, so that the phase curve $y=F(x)$ does not intersect the vertical line $x=s_i$.
It follows from this that the role of the singular points with coordinates $s_i$ is to divide the entire region of changes in the Hubble parameter $H$ into separate regions, the boundaries of which are determined by the position of the points $H=\pm s_i$ with $s_i^2$. If we renumber all the positive roots of the polynomial $P_1(H^2)$ in increasing order of $s_i^2$, we can make the following statement.

{\bf Statement}. If the initial point $H(0)$ belongs to some interval $(s_i,s_{i+1})$, then during the evolution process, the values of $H(\tau)$ will not exceed this interval.

This general statement leads to the problem of the initial value of the Hubble parameter. Since the roots $s_i^2$ are determined by the fundamental values of the parameters $\alpha_i$ for each Lovelock model, the boundaries of the intervals $(s_i,s_{i+1})$ are also fundamental. Therefore, the choice of the initial value of $H(0)$ in a particular interval is fundamental. However, the theory does not provide a physical mechanism to determine the specific choice within the universe itself. Each interval $(s_i,s_{i+1})$ corresponds to its own universe with a specific type of evolution. In addition, the intervals $(s_i,s_{i+1})$, which are bounded by singularity points, contain intervals with a constant sign of matter density. Therefore, if an interval $(s_i,s_{i+1})$ contains subintervals with negative matter density, the only available initial values for $H$ in these subintervals are those with positive matter density.

\subsection{Time to reach stationary and singular points}

 The time to reach point $H=h$ from point $h_0$, which is ``close'' to point $h$: $h\simeq h_0$, is calculated using the general formula:
    $$
        \gD\tau = \tau(h) - \tau_0 = \int\limits_{h_0}^{h} \frac{P_1(H^2)}{H^2}\frac{dH}{P_2(H^2)}
    $$
    where $\tau_0$ is the time of passage of the point $h_0$.
    Expanding the reciprocal $1/P_2(H^2)$ into a sum of partial fractions:
    $$
        \frac{1}{P_2(H^2)} = \sum\limits_{j=1}^{n-1}\frac{S_j}{H^2-h_j^2},
    $$
    where $S_j$ are some constants, we arrive at the following relation:
    $$
        \gD\tau(h,h_0)  = \int\limits_{h_0}^{h} \frac{P_1(H^2)}{H^2}\sum\limits_{j=1}^{n-1}\frac{S_j dH}{H^2-h_j^2}=
        \gD\tau_* +\int\limits_{h_0}^{h_i} \frac{C_i dH}{H^2-h^2}=
        \gD\tau_* + K_i \ln\left|\frac{h-h_i}{h_0-h_i}\right|.
    $$
    Here $\gD\tau_*$ is the value of the integral of the part of the integrand that does not contain a pole corresponding to $h=h_i$.
    Therefore, for a fixed point:
    $$
         \gD\tau(h_i,h_0)=\gD\tau_*+K \lim\limits_{h\to h_i}\ln\left|\frac{h-h_i}{h_0-h_i}\right| = \infty.
    $$
   Thus, the time to reach stationary points is infinite. This general conclusion corresponds to scenarios with an infinite expansion of the universe.
   
   For singular points, we calculate similarly:
    $$
         \gD\tau(s_i,h_0)=\gD\tau_*+K \lim\limits_{h\to s_i}\ln\left|\frac{h-h_i}{h_0-h_i}\right| = \gD\tau_*+K\ln\left|\frac{s_i-h_i}{h_0-h_i}\right| < \infty
    $$
    It follows that the time to reach singular points is finite!

\subsection{Relation to the classification of finite-time singularities}
\label{SecFTS}

It is instructive to place the singular points $H=\pm s_i$ within the standard classification of finite-time cosmological singularities introduced in \cite{NOT2005} (see also \cite{Barrow2004} and the recent review \cite{deHaro2023}). As $t \to t_s$, the four types are characterized as follows: Type I (``Big Rip''): $a\to\infty$, $\rho\to\infty$, $|p|\to\infty$; Type II (``sudden''): $a\to a_s$, $\rho\to\rho_s$, $|p|\to\infty$; Type III: $a\to a_s$, $\rho\to\infty$, $|p|\to\infty$; Type IV: $a\to a_s$, $\rho\to 0$, $|p|\to 0$, with divergences only in the higher derivatives of $H$.

The singular points studied here are reached in finite time (see the previous subsection) with a finite scale factor $a_s$, a finite Hubble parameter $H=\pm s_i$ and a finite density $\rho(s_i)$; since the matter obeys the barotropic equation of state $p=\go\rho$ with constant $\go$, the pressure remains finite as well. At the same time $|\dot H|\to\infty$, so the scalar curvature diverges and $t_s$ is a genuine curvature singularity. This combination --- finite $(a,\rho,p)$ with divergent $\dot H$ --- cannot occur in general relativity, where the field equations give $\dot H \propto -(\rho+p)$, so that a divergence of $\dot H$ requires a divergence of the pressure; that is precisely the Type II scenario of \cite{Barrow2004,NOT2005}. In the Lovelock models considered here the mechanism is different and purely geometric: the divergence of $\dot H$ originates from the vanishing of the polynomial $P_1(H^2)$ multiplying $\dot H$ in the field equation \eqref{flatF1}, i.e.\ from a degeneracy of the effective kinetic coefficient, while the matter sector remains regular. The Big Shock events introduced below may therefore be regarded as a geometric counterpart of the Type II sudden singularity (and, from the effective-fluid viewpoint of Sec.~\ref{SecEC}, exactly as a Type II singularity): they share with it the finiteness of the scale factor and of the energy density, but the role of the diverging pressure is taken over by the degeneracy of the gravitational operator itself. With respect to the matter variables $(a,\rho,p)$ alone such an event formally resembles Type IV; however, in Type IV both the density and the pressure vanish and the divergence first appears in the higher derivatives of $H$, whereas here $\rho(s_i)\neq 0$ in general and already the first derivative $\dot H$ diverges.

Singularities of this kind, arising when the coefficient in front of the highest derivative vanishes, are known in anisotropic Einstein--Gauss--Bonnet and Lovelock cosmologies as ``nonstandard'' singularities \cite{KMPT2010,Pavluchenko2024}. The present analysis provides their systematic description for the isotropic spatially flat metrics with a barotropic fluid, including the resulting scenarios in which a Big Shock replaces the Big Bang as the initial state. We also note that for the closely related sudden singularities causal geodesics are extendible and the singularity is weak in the sense of the Tipler and Kr\'olak criteria \cite{FJL2004}; since at $t_s$ both the scale factor and the density remain finite, a similar behavior can be expected for the Big Shock events, although a detailed analysis of geodesic completeness is beyond the scope of this paper.

\subsection{Ghosts in Lovelock gravity}
\label{SecGhosts}

Ghosts in the classical (Ostrogradsky) sense are absent in Lovelock gravity by construction: the Lovelock Lagrangian is the unique polynomial curvature invariant leading to field equations of strictly second order \cite{1}, so the Ostrogradsky theorem --- which implies the appearance of ghost degrees of freedom (the Ostrogradsky instability) in nondegenerate theories with higher derivatives --- does not apply to it, and the spectrum contains no additional gravitational modes with negative kinetic energy \cite{Zwiebach1985,Zumino1986}; a modern analysis of the conditions under which ghosts arise from the structure of the Lagrangian and of its constraints is given in \cite{AokiMotohashi2020}. The only ghost effect discussed in the literature for Lovelock-type theories is the Boulware--Deser ghost \cite{BoulwareDeser1985}: the graviton acquires a wrong-sign kinetic term when the theory is expanded around the ``wrong'' branch of the (A)dS vacuum. In our analysis the vacuum branches are represented by the fixed points $\pm h_i$; their dynamical stability is fully characterized by the exponents $\gl_i$, and the choice of a branch reduces to the choice of a basin of initial data, with no additional degree of freedom arising. For the models with negative couplings $\alpha_i$ (the Big Shock regime) the questions of branch selection and of the Boulware--Deser ghost require a separate analysis, which is beyond the scope of the present paper.

The regions with $\rho<0$ (phantom intervals) are likewise unrelated to ghosts in the matter sector. The matter in our model is an ordinary barotropic fluid with $p=\go\rho$; the negativity of the density in these intervals means that it is the geometry --- the higher-order Lovelock terms --- that requires $\rho<0$ for the consistency of the dynamical equations: a normal fluid cannot support the cosmological evolution with such values of the Hubble parameter. At the same time, the phase trajectories with $\rho>0$ never penetrate into the phantom intervals: their boundaries are the fixed points $\pm h_i$, which are reached only asymptotically (see also Appendix~B: the phantom intervals require negative couplings $\alpha_i$ and never overlap with the central basin). Thus, for a universe filled with ordinary matter the corresponding ranges of $H$ turn out to be ``forbidden'', which in principle opens a possibility of an observational test: realization of the values of $H$ from such ranges would require matter of a phantom type \cite{C02,GPRS06}. We also note that the extendibility of solutions through critical points of polynomial origin in gravity with maximally extended Gauss--Bonnet terms was established already in \cite{KitauraWheeler1991}, in agreement with the ``mild'' character of our singular barriers.

\section{Examples of cosmological dynamics analysis}

We will illustrate the general qualitative analysis with several simple examples.

\subsection{General relativity}
Lovelock gravity reduces to general relativity when $n=1$ and $N=4$, in which case \eqref{dotH} becomes:
\begin{equation}
	\frac{dH}{d\tau} = -\frac{3}{2}\varepsilon H^2.
	\label{HGR}
\end{equation}

\begin{figure}
    \centering
    \includegraphics[width=0.8\textwidth]{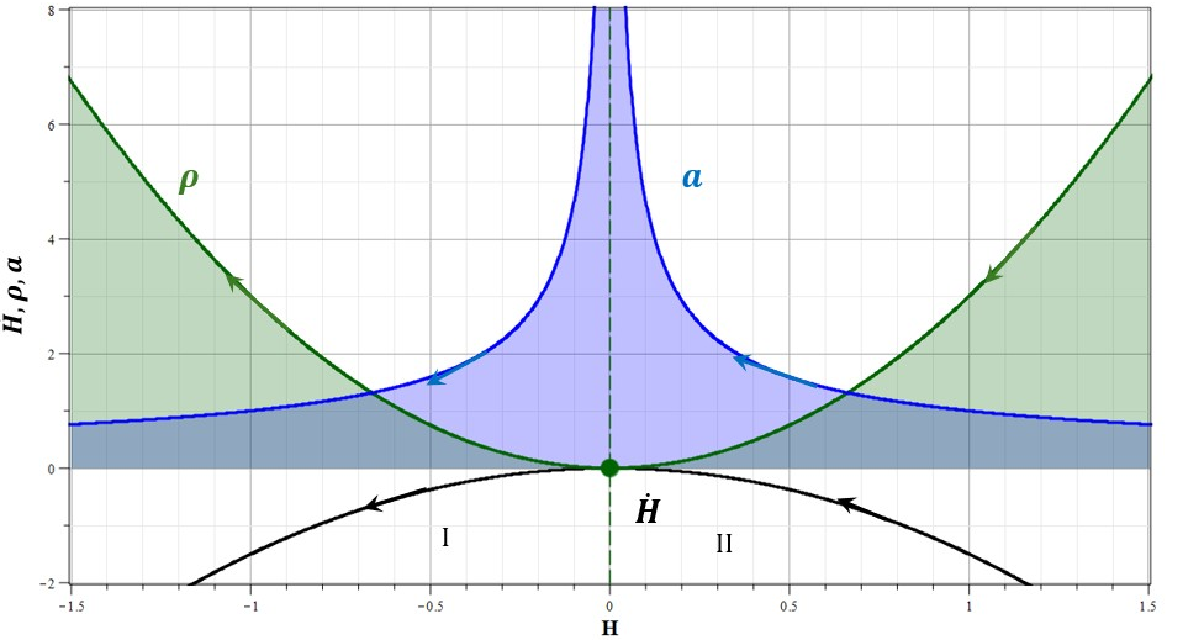}\\
    \textbf{a} \\ \vspace{0.2cm}
    \includegraphics[width=0.8\textwidth]{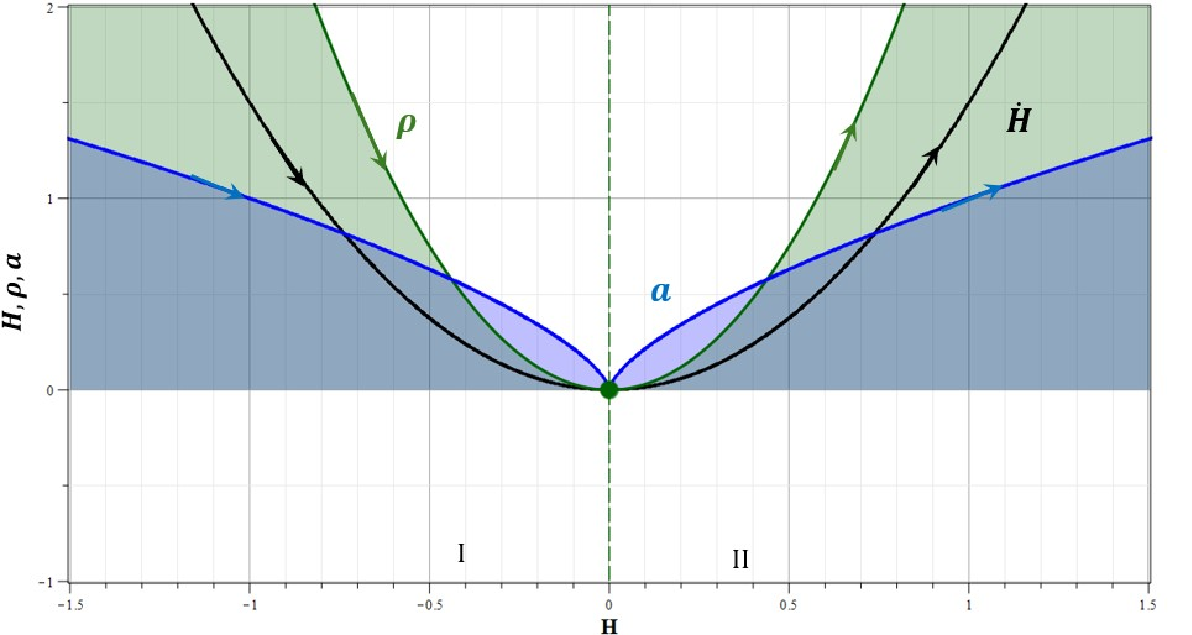}\\
    \textbf{b}\\
    \caption{Phase portraits of the system \eqref{HGR} together with the dependencies $\rho=\rho(H)$ and $a=a(H)$: a) $\varepsilon=+1$, b) $\varepsilon=-1$ }
    \label{Fig1}
    \end{figure}

The phase portrait of the system \eqref{HGR} is presented in Fig. \ref{Fig1}ab. Fig. \ref{Fig1}a shows the case with $\varepsilon =+1$, while Fig. \ref{Fig1}b shows the case with $\varepsilon =-1$. The blue dotted line represents the graph of the matter density as a function of $H$. The case of a constant $\omega = -1$, which is equivalent to $\dot{H}=0$, corresponds to a quasi-vacuum state with a de Sitter evolution of the scale factor $a\sim e^{H_0 t}$.

In all cases (except for $\omega=-1$) with a sign-constant $\Gamma(t)$, the system \eqref{HGR} has one special point, the degenerate fixed point $H=0$. In Figures \ref{Fig1}ab, the arrows indicate the direction of parameter change during evolution.

In both cases, $\varepsilon=\pm 1$, there are two different scenarios for the evolution of the universe. In Fig. \ref{Fig1}ab, these scenarios are denoted by the numbers $I$ and $II$.
For all scenarios, the point $H=0$ is attractive on one side and repulsive on the other. In the case of $\veps=+1$, scenario $I$ describes the contraction of the universe $a\to 0$ as $t\to \infty$, with its birth from a state with $a=\infty$ and $\rho=0$. Scenario $II$ of the same sign $\veps$ describes the standard Friedmann evolution scenario (Big Bang) from a state with $a=0$ and $\rho=\infty$ to a state with $a=\infty$ and $\rho=0$. In contrast, the scenarios with $\veps=-1$ are paradoxical. Scenario $I$ describes the birth of the universe from the state $a=\infty$ and $\rho=\infty$ and its transition to the state $a=0$ and $\rho=0$. Scenario $II$ describes the birth of the universe from the state $a=0$ and $\rho=0$ and its transition to the state $a=\infty$ and $\rho=\infty$.

   \begin{center}
\begin{tabular}{|c|c|c|}
\hline
 Parameter  & \multicolumn{2}{|c|}{Scenarios, $\veps=+1$} \\
    \hline
  Number   &  $ I $ & $II$  \\
    \hline
 Interval & $(-\infty,h_0)$ & $(h_0,\infty)$\\
       \hline
 $\rho$ & $ \infty\leftarrow 0$  & $0\leftarrow\infty $ \\
       \hline
 $a$ & $0 \leftarrow \infty $ &  $\infty \leftarrow 0 $\\
       \hline
 $H$ & $-\infty \leftarrow 0$ &  $0 \leftarrow \infty$\\
       \hline
 $\dot{H} $ & $-\infty \leftarrow 0$ &  $ 0\leftarrow -\infty$\\
       \hline
\end{tabular}
\begin{tabular}{|c|c|c|}
\hline
 Parameter  & \multicolumn{2}{|c|}{Scenarios, $\veps=-1$} \\
    \hline
  Number   &  $ I $ & $II$  \\
    \hline
 Interval & $(-\infty,h_0)$ & $(h_0,\infty)$\\
       \hline
 $\rho$ & $ \infty \rightarrow 0$  & $0 \rightarrow \infty  $ \\
       \hline
 $a$ & $ \infty \rightarrow 0 $ &  $ 0 \rightarrow \infty$\\
       \hline
 $H$ & $-\infty \rightarrow 0$ &  $0 \rightarrow \infty $\\
       \hline
 $\dot{H} $ & $\infty \rightarrow 0$ &  $ 0\rightarrow \infty$\\
       \hline
\end{tabular}
\vspace{0.15cm}
\

Table 1. \label{Tab1} Possible evolution scenarios in \rf{HGR} models. In each cell the left/right entries give the limiting values of the quantity at the left/right end of the interval; the arrow indicates the direction of time evolution.
\end{center}

Equation \eqref{HGR} has an exact solution:
$$
     H = \frac{2\varepsilon}{3 (\tau-\tau_0)},
$$
where $\tau_0$ is an arbitrary integration constant that determines the position of the cosmological singularity $H$ in time. Based on this general solution, we find:
$$
     a = a_0 |\tau-\tau_0|^{2\varepsilon /3}.
$$
It can be seen that for $\varepsilon=+1$, $\lim\limits_{\tau \to\infty}a(\tau)=\infty$ --- the Universe expands according to the Friedmann scenario, while for $\varepsilon=-1$, $\lim\limits_{\tau\to\infty}a(\tau)=0$ --- the Universe contracts, while the density paradoxically tends to zero.

This rather detailed analysis is presented here because the special point $H=0$ exists for all Lovelock models discussed in the following sections. However, for other models, the behavior of the universe at this point will be identical to that described in this section.

\subsection{Einstein-Gauss-Bonnet gravity}
\label{SecEGB}

When $n=2$ and $N=5$, Lovelock gravity reduces to Einstein-Gauss-Bonnet gravity. In this case, \eqref{dotH} becomes:
\begin{equation}
	\frac{dH}{d\tau} = -\varepsilon H^2\frac{6 + 12\alpha_2H^2}{3+12\alpha_2 H^2}
	\label{HEGB}
\end{equation}
The model corresponding to \rf{HEGB} has a singular point $H_0=0$ in the case $\alpha_2 < 0$, two fixed points
$$
h_1 = \pm \sqrt{\frac{1}{2|\alpha_2|}}
$$
and two singular special points
$$
    s_1 = \pm \sqrt{\frac{1}{4|\alpha_2|}}.
$$

In the case $\alpha_2 > 0$, there is a single singular point $H_0=0$. In this case, the phase curves of the system are similar to the phase curves for the model \eqref{HGR}, which are shown in Fig. \ref{Fig1}ab.

\begin{figure}
    \centering
    \includegraphics[width=0.8\textwidth]{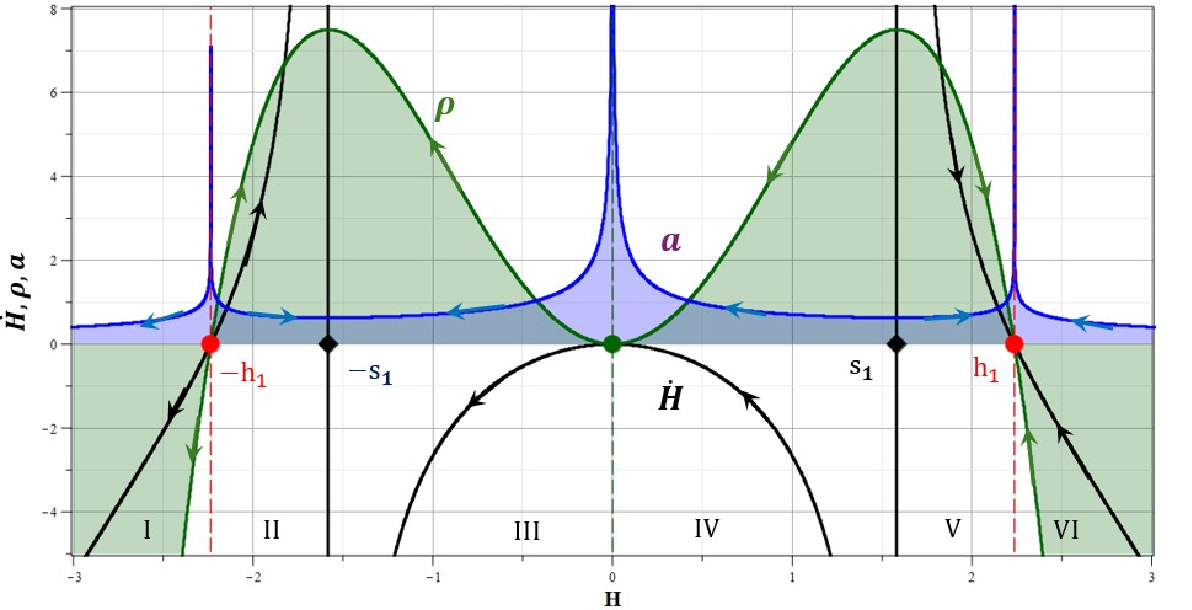}\\
    \textbf{a} \\ \vspace{0.2cm}
    \includegraphics[width=0.8\textwidth]{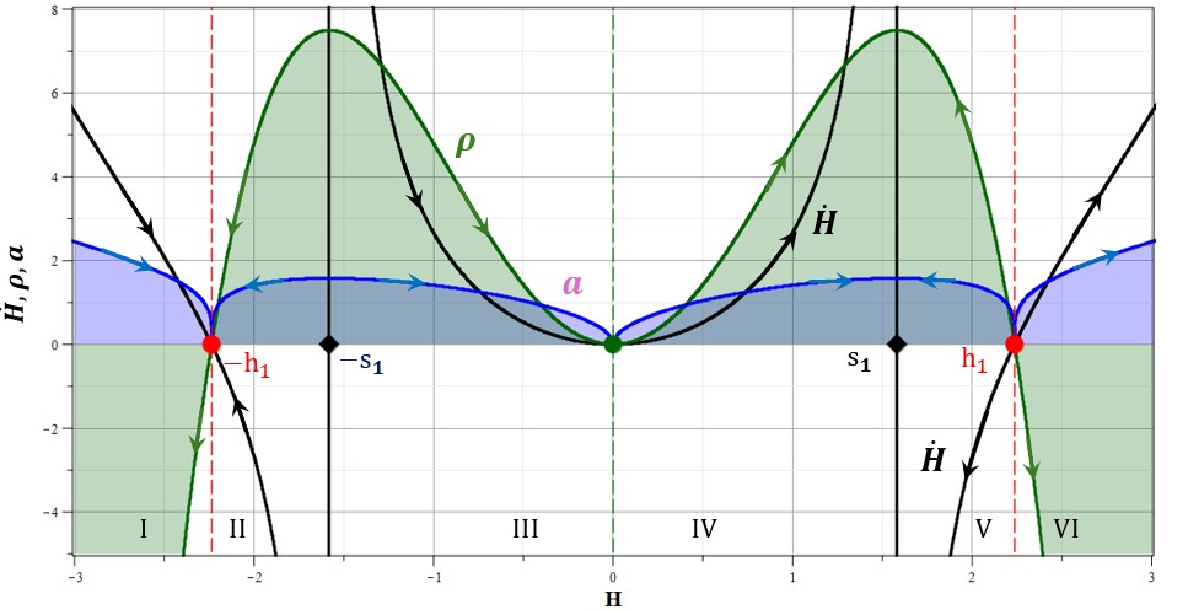}\\
    \textbf{b}
    \caption{Phase portraits of the \rf{HEGB} system for the case $\alpha_2 = -0.1$, $|\go|=1/3$: a) $\varepsilon=+1$, b) $\varepsilon=-1$ }
    \label{Fig2}
    \end{figure}

Fig. \ref{Fig2}ab shows the phase diagrams of the system \eqref{HEGB} and the dependence of $\rho$ and $a$ on $H$ in the case of $\alpha_2=-0.1$ for both cases of $\varepsilon = +1$ (a) and $\varepsilon = -1$ (b). The physically allowed intervals with positive density of matter are as follows: $(-h_1,-s_1),~(-s_1,0),~(0,s_1),~(s_1,h_1)$. When $H > h_1$ and $H < -h_1$, the density is negative. Note that in intervals with $\rho>0$, the density remains finite even at singular points where $\dot{H}$ goes to infinity.

    \subsubsection{Possible evolution scenarios}

    The considered variant of the Einstein-Gauss-Bonnet model with $n=2$ and $N=5$, corresponding to the specific value $\alpha_2=-0.1$, represents the most general variant of models of this type. When $\alpha_2$ is changed in the region of negative values, the general differences in phase diagrams and the evolution of $\rho$ and $a$ will be reduced to changes in the position of the fixed and the singular points, while maintaining one special point $H=0$. For $\alpha_2=-0.1$ the corresponding points are $s_1\approx 1.581$ and $h_1\approx 2.236$.

    All scenarios in Fig. \ref{Fig2}ab are numbered with Roman numerals $I,II,\ldots,VI$, and their main properties are collected in Tables 2a and 2b, analogously to the \rf{HGR} models.
    Scenarios $III$ and $IV$: $(-s_1,0),~(0,s_1)$, limited by the point $H=0$ and singular points, in the case of $\varepsilon=+1$ are hardly distinguishable from the standard scenario of Friedmann evolution, with the difference that the initial point of evolution does not start from the cosmological singularity, but from a state with finite values of the matter density and scale factor, as well as with a finite positive Hubble parameter. The final values of $H=\pm s_1$ correspond to the initial de Sitter and anti-de Sitter evolution regimes, but with an infinite initial value of $\dot{H}$. The $IV$ scenario differs significantly from the Big Bang scenario, in which the universe is born from a state with infinite density and zero scale factor. In both the Big Bang scenario and scenario $IV$ the final point of evolution is a universe with $a=\infty$ and zero density, which is approached according to the Friedmann scenario, the time needed to reach the state with $\rho=0$ being infinite. However, since in scenario $IV$ the initial values of the parameters differ radically from those of the Big Bang scenario, this scenario can be referred to as the ``Big Shock''. It corresponds to a situation where a universe with a finite density and scale factor suddenly experiences an initial (infinite) acceleration and then expands with a deceleration. Similarly, scenario $III$ differs from scenario $I$ in classical models \rf{HGR}.

    Scenario $V$: $(s_1,h_1)$, in the case of $\varepsilon=+1$, describes the ``birth'' of the Universe from a state with a finite density and a finite scale factor, and its subsequent transition to a de Sitter expansion with a zero density. This scenario is similar to scenario $IV$, with the difference that the asymptotic limit is not a Friedmann scenario, but a de Sitter scenario, and that the initial value of $\dot{H}$ is positive. Therefore this scenario can also be assigned to the Big Shock type. Since the signs of the initial values of $\dot{H}$ are different in scenarios $IV$ and $V$, the former may be called a negative Big Shock and the latter a positive Big Shock.
    
    Similarly, scenario $II$ with $\varepsilon=+1$ is similar to scenario $III$ and describes the birth of the universe from a state with zero density and an infinite scale factor, which, in the form of an anti-de Sitter contraction, transitions to a state with a finite matter density but an infinite positive rate of change in $H$. The final state, a universe with a finite scale factor and density, is reached from any state with $H<-h_1$ in a finite time.
    
    Scenarios $I$ and $VI$ are physically unrealizable in the standard interpretation, since they describe a universe with a negative matter density. However, if $p$ and $\rho$ are regarded as the effective pressure and density of a phantom scalar field \cite{C02,GPRS06}, such an approach may be useful in models with matter in the form of scalar fields.
    
    All scenarios $I-VI$ corresponding to $\veps=-1$ can be analyzed in a similar way to the scenarios with $\veps=+1$. However, as in the case of \rf{HGR} models, they have a number of paradoxical properties that make them unsuitable for comparison with the properties of the observed universe.

    \begin{center}
\begin{tabular}{|c|c|c|c|c|c|c|}
\hline
 Parameter  & \multicolumn{6}{|c|}{Scenarios $\veps=+1$} \\
\hline
  Number   &  $ I $ & $II$ & $III$ & $IV$ & $V$ & $VI$ \\
    \hline
Interval & $(-\infty,-h_1)$  & $(-h_1,-s_1)$  & $(-s_1,0)$ & $(0,s_1)$ & $(s_1,h_1)$ & $(h_1,\infty)$\\
       \hline
 $\rho$ & $-\infty \leftarrow 0 $  &  $0 \rightarrow \rho(-s_1)$ &  $ \rho(-s_1)\leftarrow 0 $ & $ 0 \leftarrow \rho(s_1)$  & $\rho(s_1) \rightarrow 0$ & $0 \leftarrow -\infty$\\
       \hline
 $a$ & $ 0\leftarrow \infty$ &  $\infty \rightarrow a(-s_1)$ & $ a(-s_1)\leftarrow \infty$ & $\infty\leftarrow a(s_1)$  & $a(s_1)\rightarrow \infty$ & $\infty \leftarrow 0$ \\
       \hline
 $H$ & $-\infty\leftarrow -h_1$   &  $-h_1\rightarrow -s_1$ & $-s_1 \leftarrow 0 $ & $0 \leftarrow s_1 $ & $s_1\rightarrow h_1$ & $h_1 \leftarrow \infty$ \\
       \hline
 $\dot{H}$ & $-\infty\leftarrow 0$   &  $0\rightarrow \infty$ & $-\infty \leftarrow 0$ & $0\leftarrow -\infty$ & $\infty \rightarrow 0$ & $0\leftarrow -\infty$ \\
       \hline
\end{tabular}
\vspace{0.15cm}
\

Table 2a \label{Tab2a} Evolution scenarios in \rf{HEGB} models with $\veps=+1$.
\end{center}

  \begin{center}
\begin{tabular}{|c|c|c|c|c|c|c|}
\hline
 Parameter  & \multicolumn{6}{|c|}{Scenarios $\veps=-1$} \\
\hline
  Number   &  $ I $ & $II$ & $III$ & $IV$ & $V$ & $VI$ \\
    \hline
Interval & $(-\infty,-h_1)$  & $(-h_1,-s_1)$  & $(-s_1,0)$ & $(0,s_1)$ & $(s_1,h_1)$ & $(h_1,\infty)$\\
       \hline
 $\rho$ & $\infty \rightarrow 0 $  &  $0\leftarrow \rho(-s_1)$ &  $\rho(-s_1) \rightarrow 0$ & $ 0 \rightarrow \rho(s_1)$  & $ \rho(s_1) \leftarrow 0$ & $ 0\rightarrow -\infty$ \\
       \hline
 $a$ & $\infty \rightarrow 0 $ &  $0 \leftarrow a(-s_1)$ & $ a(-s_1) \rightarrow 0 $ & $ 0 \rightarrow a(s_1) $  & $ a(s_1)\leftarrow 0$ & $0 \rightarrow \infty$  \\
       \hline
 $H$ & $-\infty\rightarrow -h_1$   &  $-h_1\leftarrow -s_1$ & $-s_1 \rightarrow 0$ & $0\rightarrow s_1$ & $s_1\leftarrow h_1$ & $h_1\rightarrow +\infty$ \\
       \hline
 $\dot{H}$ & $\infty\rightarrow 0$   &  $0\leftarrow -\infty$ & $\infty \rightarrow 0$ & $0\rightarrow \infty$ & $-\infty \leftarrow 0$ & $0\rightarrow +\infty$ \\
       \hline
\end{tabular}
\vspace{0.15cm}
\

Table 2b \label{Tab2b} Evolution scenarios in the \rf{HEGB} models when $\veps=-1$.
\end{center}

\begin{figure}
    \centering
    \includegraphics[width=0.8\textwidth]{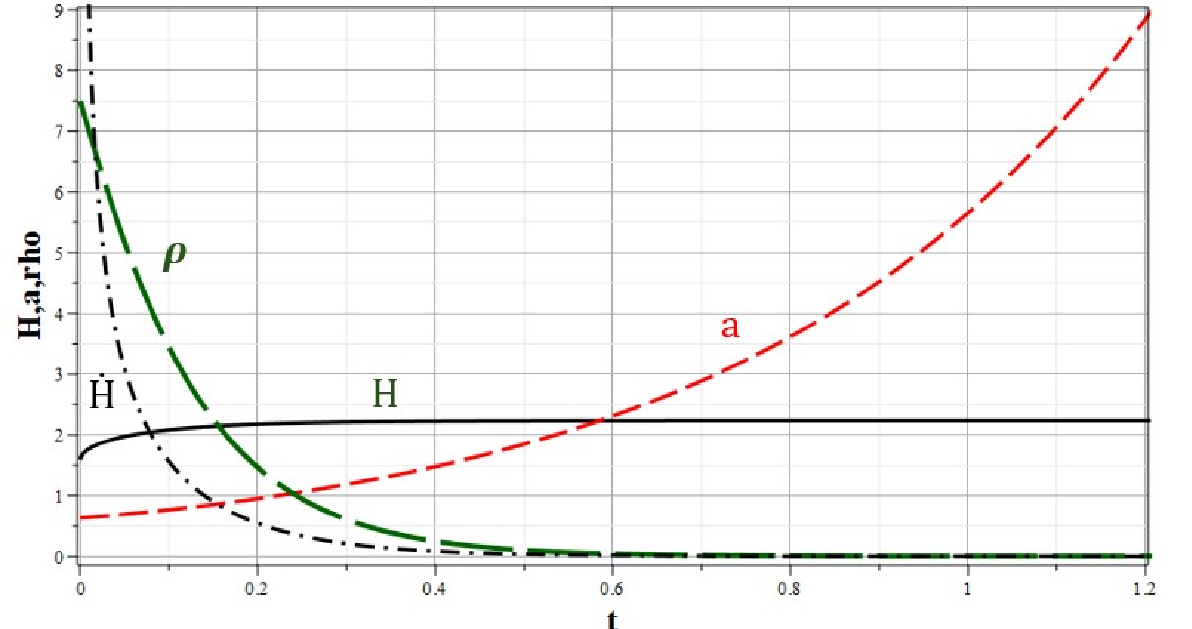}\\
    \textbf{a} \\ \vspace{0.2cm}
    \includegraphics[width=0.8\textwidth]{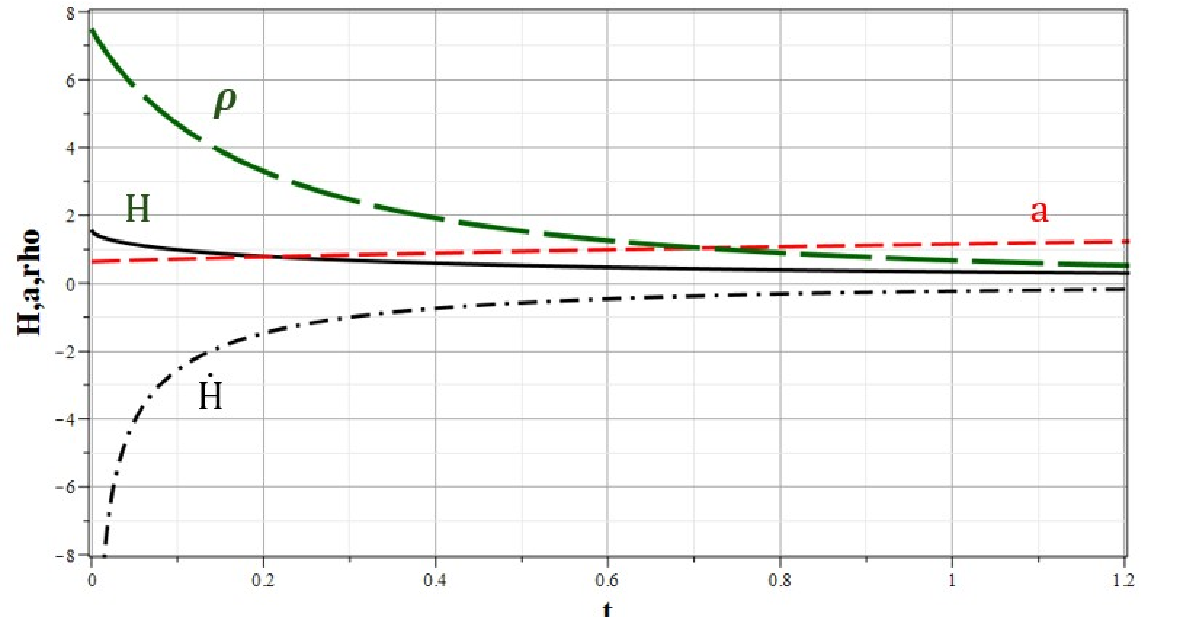}\\
    \textbf{b}
    \caption{Time evolution of $\dot{H},~H,~a,~\rho$ for the model \rf{HEGB} with the parameters of Fig.~\ref{Fig2} and initial conditions: a) $H_0=1.59$, b) $H_0=1.58$ }
    \label{Fig21}
    \end{figure}

Fig. \ref{Fig21} shows the time evolution of $\dot{H},~H,~a$ and $\rho$ for the values of the parameters $\ga_i$ corresponding to the phase diagram in Fig. \ref{Fig2}a. The initial values $H(0)=H_0$ are chosen so that in the case a) the evolution follows the Big Shock scenario $V$, and in the case b) the scenario $IV$. The initial values of $a(t)$ were computed analytically from the formula \rf{Sola}. A singular point of the model lies between $H_0=1.58$ and $H_0=1.59$. The plots illustrate the characteristic features of the evolution in the corresponding scenarios. The Big Shock scenario $V$ is close to the de Sitter scenario, deviating from it near the initial moment of time, when $\dot{H}$ takes large positive values; for $t>0.1$ the scale factor $a$ varies almost exponentially. Scenario $IV$ displays all the features of the Friedmann scenarios, with $a$ changing as a power law in time.

\subsection{Cubic Lovelock gravity}
\label{SecCubic}

\subsubsection{Equations and singular points of the model}

To illustrate the possible types of evolution in general, consider the case of $N=7$ and $n=3$. In this case, the model equations become:
\begin{equation}
	\frac{dH}{d\tau} = -\varepsilon H^2\frac{15 + 180\alpha_2 H^2 + 360\alpha_3 H^4}{5 + 120\alpha_2H^2 + 360\alpha_3 H^4}.
	\label{H37}
\end{equation}
The roots of the polynomial $P_1(z)$ are as follows:
\begin{equation}
	S_{1,2}^2 =- \frac{1}{12\alpha_3}\left(2\alpha_2\pm\sqrt{4\alpha_2^2-2\alpha_3}\right),
	\label{SH37}
\end{equation}
Accordingly, the roots of the polynomial $P_2(z)$ are as follows:
\begin{equation}
	H_{1,2}^2 =- \frac{1}{12\alpha_3}\left(3\alpha_2\pm\sqrt{9\alpha_2^2-6\alpha_3}\right).
	\label{ZH37}
\end{equation}
It follows that $P_1(z)$ has real roots for $\alpha_3 \le 2\alpha_2^2$, and $P_2(z)$ for $\alpha_3 \le 3\alpha_2^2/2$. At $\alpha_3 = 3\alpha_2^2/2$ the roots of $P_2(z)$ merge, $h_1^2 =h_2^2 = -1/(6\alpha_2)$, while the roots of $P_1(z)$ are $s_1^2=-1/(18\alpha_2)$ and $s_2^2=-1/(6\alpha_2)$, so that the larger root of $P_1$ coincides with the degenerate root of $P_2$: on this curve in the parameter plane the singular barrier degenerates into a fixed point, in agreement with the general statement of Appendix~B. Thus, as the parameters $\alpha_2$ and $\alpha_3$ approach this boundary from below, the dynamics of \rf{H37} degenerates into that of the model \eqref{HEGB}.

\begin{figure}
    \centering
    \includegraphics[width=0.44\textwidth]{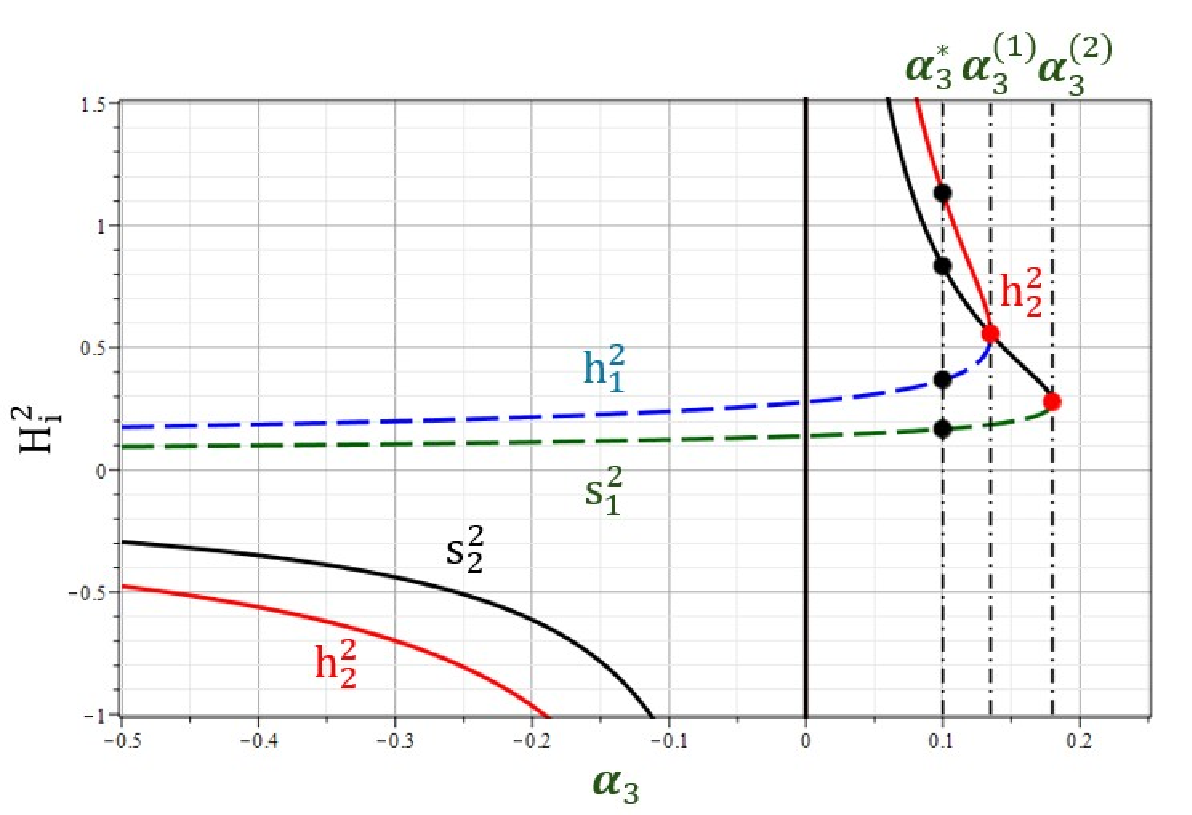}
    \includegraphics[width=0.45\textwidth]{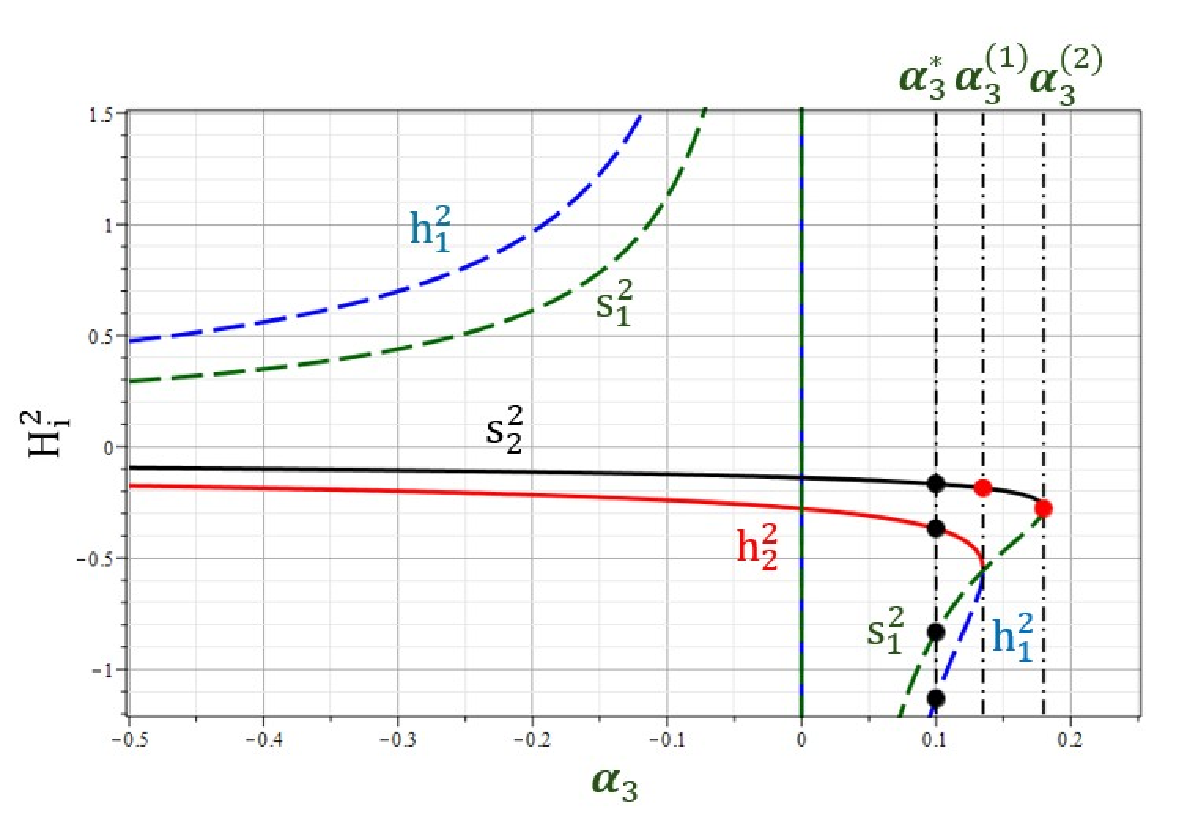}\\
    a \hspace{7cm} b
    \caption{Roots of the polynomials as functions of $\ga_3$: a) $\ga_2=-0.3$, b) $\ga_2=0.3$ }
    \label{FigR}
    \end{figure}

\begin{figure}
    \centering
    \includegraphics[width=0.8\textwidth]{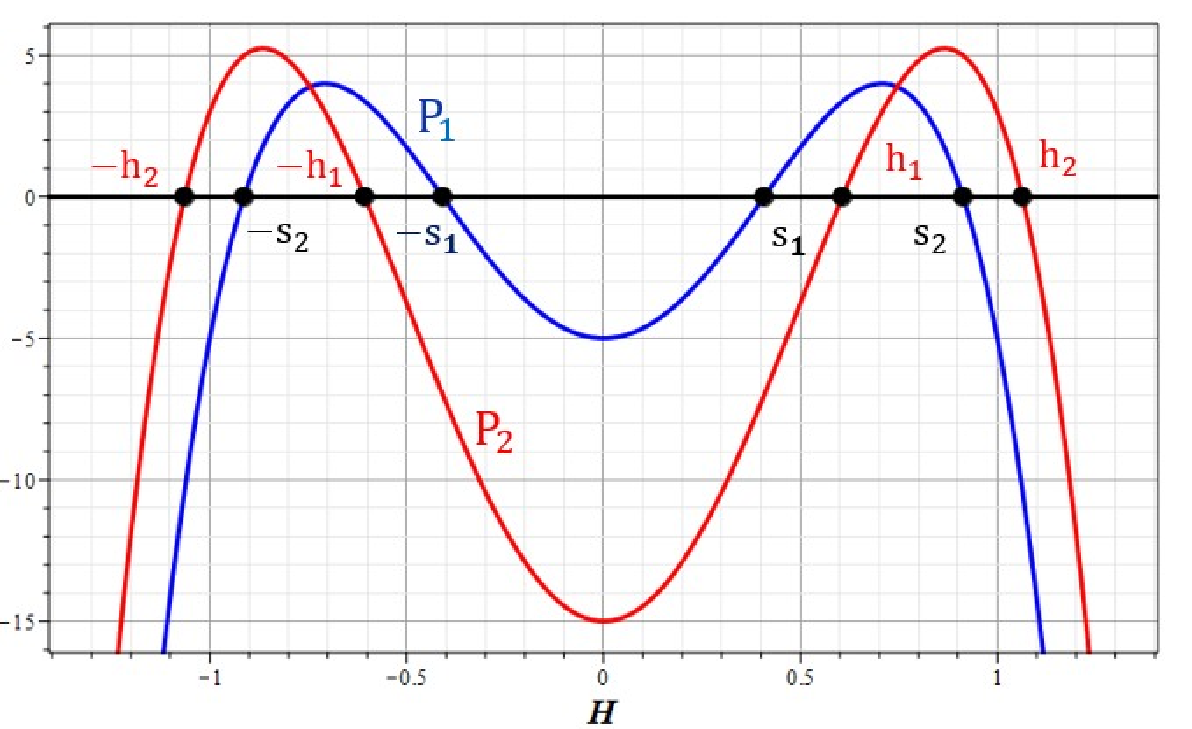}
    \caption{The polynomials $-P_1(z)$ and $-P_2(z)$ as functions of $H$ for $\ga_2=-0.3$ and $\ga_3=0.1$ }
    \label{FigP}
    \end{figure}

For both polynomials to have positive roots, additional conditions are required. The analysis shows that the roots of the two polynomials can be positive simultaneously if $\ga_2<0$. Fig. \ref{FigR}a shows the roots of both polynomials as functions of $\ga_3$ for $\ga_2=-0.3$, and Fig. \ref{FigR}b shows the same dependencies for $\ga_2=0.3$. In Fig. \ref{FigR}a,b the value $\ga_3^*=0.1$ is the value of $\ga_3$ for which the phase portrait of the model \eqref{H37} is constructed, while $\ga_3^{(1)}$ and $\ga_3^{(2)}$ are the values of $\alpha_3$ at which the roots of $P_2(z)$ and $P_1(z)$, respectively, merge. In the case $\ga_2>0$ each of the polynomials has a single positive root, which reduces the general character of the phase portrait to that of the model \eqref{HEGB}. To illustrate the general relation between the polynomials and their roots, Fig. \ref{FigP} shows the graphs of $-P_1(z)$ and $-P_2(z)$ for $\ga_2=-0.3$ and $\ga_3=0.1$.

\subsubsection{Phase diagrams}

\begin{figure}
    \centering
    \includegraphics[width=0.8\textwidth]{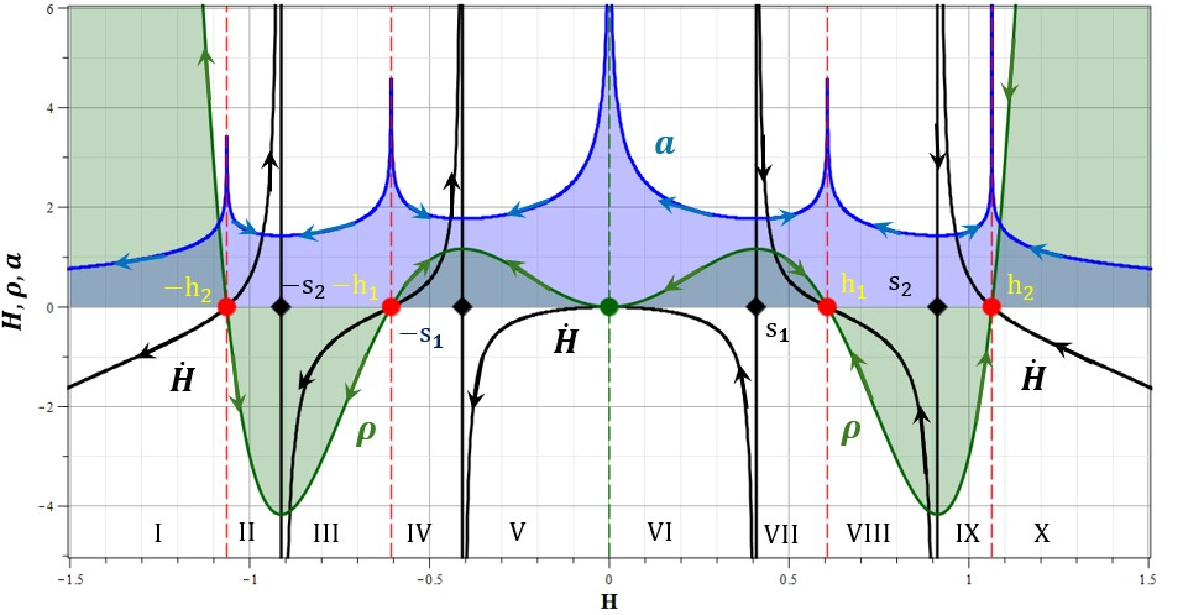}\\
    \textbf{a} \\ \vspace{0.2cm}
    \includegraphics[width=0.8\textwidth]{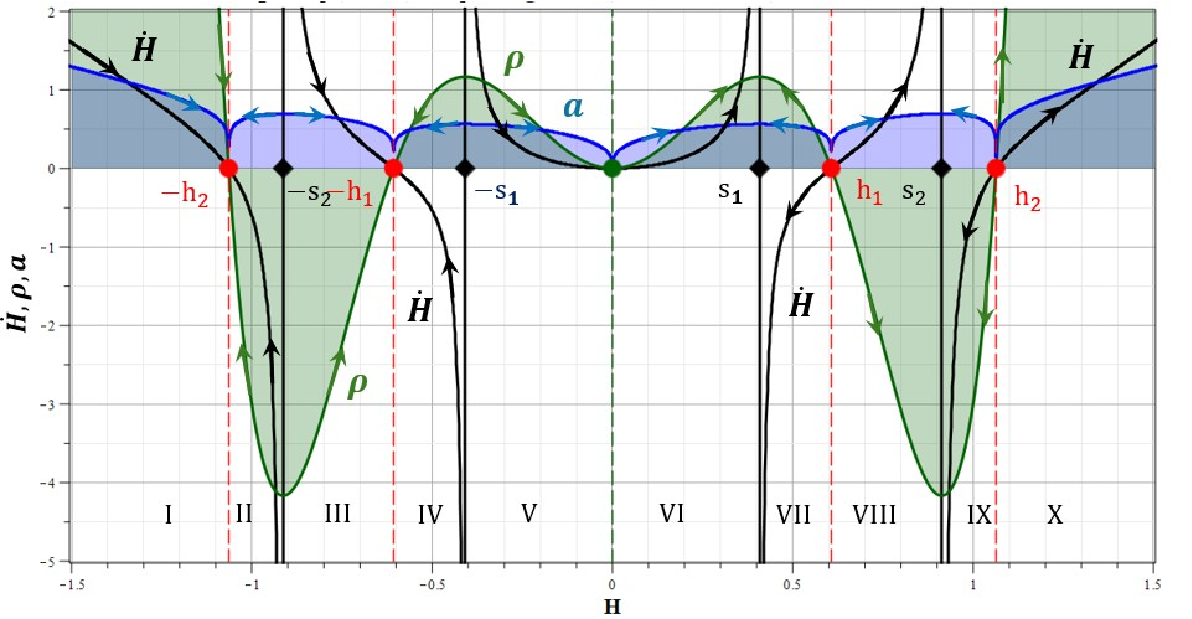}\\
    \textbf{b}
    \caption{Phase portraits of the system \eqref{H37} for the case $\alpha_{2}=-0.3,~\alpha_{3}=0.1$: a) $\varepsilon=+1$, b) $\varepsilon=-1$ }
    \label{Fig3}
    \end{figure}

Fig. \ref{Fig3}ab shows the phase diagrams of the system \rf{H37} for the situations $\varepsilon = +1$ (a) and $\varepsilon = -1$ (b) at $\alpha_{2}=-0.3,~\alpha_{3}=0.1$. The values of the coefficients $\alpha_{2},~\alpha_{3}$ are chosen so that the polynomials $P_1(z)$ and $P_2(z)$ have two real positive roots. The variants with one real root are analogous to models \eqref{HEGB}, and the variants with no positive real roots are analogous to models \eqref{HGR}. For both signs of $\veps$, the physically unrealizable regions with $\rho(H)<0$ form two bounded intervals $(-h_2,-h_1)$ and $(h_1,h_2)$, inside which the singular separatrix points $\pm s_2$ are located. As in the case of model \eqref{HEGB}, the density remains finite in all intervals with finite values of $H$, even at the singular points of $\dot{H}$.

    \subsubsection{Possible evolution scenarios}

    The entire interval of $H$ values is divided into ten subintervals in Figures \ref{Fig3}ab,
    corresponding to ten different types of evolution scenarios, which mostly repeat the scenarios corresponding to the models \rf{HEGB}. Tables 3a and 3b present all the main characteristics of the possible scenarios for the model \rf{H37} with $\alpha_2=-0.3 $ and $\alpha_3=0.1$.
     \begin{center}
\begin{tabular}{|c|c|c|c|c|c|}
\hline
 Parameter  & \multicolumn{5}{|c|}{Scenarios $\veps=+1$} \\
\hline
  Number   &  $ I $ & $II$ & $III$ & $IV$ & $V$\\
    \hline
Interval & $(-\infty,-h_2)$  & $(-h_2,-s_2)$  & $(-s_2,-h_1)$ & $(-h_1,-s_1)$ & $(-s_1,0)$\\
       \hline
 $\rho $ & $ \infty \leftarrow 0 $  &  $0 \rightarrow\rho(-s_2) $ &  $\rho(-s_2) \leftarrow 0$ & $ 0\rightarrow \rho(-s_1)$  & $ \rho(-s_1) \leftarrow 0$ \\
       \hline
 $a$ & $ 0 \leftarrow \infty $ &  $\infty \rightarrow a(-s_2)$ & $ a(-s_2) \leftarrow\infty$ & $\infty \rightarrow a(-s_1)$  & $a(-s_1) \leftarrow \infty $  \\
       \hline
 $H$ & $-\infty\leftarrow -h_2$   &  $-h_2\rightarrow -s_2$ & $-s_2 \leftarrow -h_1$ & $-h_{1}\rightarrow -s_1$ & $ -s_1 \leftarrow 0$ \\
       \hline
 $\dot{H}$ & $-\infty\leftarrow 0$   &  $0\rightarrow \infty$ & $-\infty \leftarrow 0$ & $0\rightarrow \infty$ & $ -\infty \leftarrow 0$ \\
       \hline
\end{tabular}

\begin{tabular}{|c|c|c|c|c|c|}
\hline
 Parameter  & \multicolumn{5}{|c|}{Scenarios $\veps=+1$} \\
\hline
  Number   &  $ VI $ & $VII$ & $VIII$ & $IX$ & $X$\\
    \hline
Interval & $(0,s_1)$  & $(s_1,h_1)$  & $(h_1,s_2)$ & $(s_2,h_2)$ & $(h_2,\infty)$\\
       \hline
 $\rho$ & $ 0\leftarrow \rho(s_1)$  &  $\rho(s_1) \rightarrow 0$ &  $0 \leftarrow \rho(s_2)$ & $ \rho(s_2)\rightarrow 0 $  & $ 0 \leftarrow\infty$ \\
       \hline
 $a$ & $\infty \leftarrow a(s_1)$ &  $a(s_1)\rightarrow \infty$ & $ \infty \leftarrow a(s_2)$ & $a(s_2) \rightarrow \infty$  & $\infty \leftarrow 0$ \\
       \hline
 $H$ & $0 \leftarrow s_1$   &  $s_1\rightarrow h_1$ & $h_1 \leftarrow s_2$ & $s_2\rightarrow h_2$ & $h_2 \leftarrow\infty $ \\
       \hline
 $\dot{H}$ & $0 \leftarrow -\infty$   &  $\infty\rightarrow 0$ & $ 0 \leftarrow-\infty $ & $ \infty \rightarrow 0$ & $ 0\leftarrow -\infty$ \\
       \hline
\end{tabular}
\vspace{0.15cm}
\

Table 3a. \label{Tab3a} Evolution scenarios in \rf{H37} models with $\veps=+1$.
\end{center}

 \begin{center}
\begin{tabular}{|c|c|c|c|c|c|}
\hline
 Parameter  & \multicolumn{5}{|c|}{Scenarios $\veps=-1$} \\
\hline
  Number   &  $ I $ & $II$ & $III$ & $IV$ & $V$\\
    \hline
Interval & $(-\infty,-h_2)$  & $(-h_2,-s_2)$  & $(-s_2,-h_1)$ & $(-h_1,-s_1)$ & $(-s_1,0)$\\
       \hline
 $\rho$ & $\infty \rightarrow 0 $  &  $0\leftarrow \rho(-s_2)$ &  $\rho(-s_2) \rightarrow 0$ & $ 0 \leftarrow\rho(-s_1)$  & $ \rho(-s_1) \rightarrow 0$ \\
       \hline
 $a$ & $\infty \rightarrow 0 $ &  $0 \leftarrow a(-s_2)$ & $ a(-s_2) \rightarrow 0 $ & $ 0 \leftarrow a(-s_1) $  & $ a(-s_1)\rightarrow 0$  \\
       \hline
 $H$ & $-\infty\rightarrow -h_2$   &  $-h_2\leftarrow -s_2$ & $-s_2 \rightarrow -h_1$ & $-h_{1}\leftarrow -s_1$ & $-s_1\rightarrow 0$ \\
       \hline
 $\dot{H}$ & $\infty\rightarrow 0$   &  $0\leftarrow -\infty$ & $\infty \rightarrow 0$ & $0\leftarrow -\infty$ & $ \infty \rightarrow 0$ \\
       \hline
\end{tabular}

\begin{tabular}{|c|c|c|c|c|c|}
\hline
 Parameter  & \multicolumn{5}{|c|}{Scenarios $\veps=-1$} \\
\hline
  Number   &  $ VI $ & $VII$ & $VIII$ & $IX$ & $X$\\
    \hline
Interval & $(0,s_1)$  & $(s_1,h_1)$  & $(h_1,s_2)$ & $(s_2,h_2)$ & $(h_2,\infty)$\\
       \hline
 $\rho$ & $0 \rightarrow \rho(s_1)$  &  $ \rho(s_1)\leftarrow 0$ &  $0 \rightarrow \rho(s_2)$ & $ \rho(s_2)\leftarrow 0$  & $ 0\rightarrow\infty$ \\
       \hline
 $a$ & $0 \rightarrow a(s_1)$ &  $ a(s_1)\leftarrow 0$ & $ 0\rightarrow a(s_2)$ & $ a(s_2)\leftarrow 0$  & $ 0 \rightarrow \infty$ \\
       \hline
 $H$ & $0 \rightarrow s_1 $   &  $s_1\leftarrow h_1$ & $h_1 \rightarrow s_2$ & $s_2\leftarrow h_2$ & $h_2\rightarrow\infty$ \\
       \hline
 $\dot{H}$ & $0 \rightarrow \infty $   &  $-\infty  \leftarrow 0 $ & $ 0 \rightarrow \infty$ & $ -\infty\leftarrow 0$ & $ 0 \rightarrow \infty$ \\
       \hline
\end{tabular}
\vspace{0.15cm}
\

Table 3b. \label{Tab3b} Evolution scenarios in the \rf{H37} models with $\veps=-1$.
\end{center}

As in the models \rf{HEGB} in the case $\veps=+1$ in the model \rf{H37}, there are ``Big Shock'' scenarios ($VI,VII$) in which the initial state of the universe has a finite density and a finite scale factor, but an infinite value of $\dot{H}$, of opposite signs in the two scenarios. At the end of the evolution, in both scenarios the universe transitions to a state with zero density and an infinite scale factor. Additional types are the scenarios $VIII$ and $IX$, describing the birth of the universe from a state with a finite scale factor and a finite matter density, but with a negative value of the latter. The final point of evolution in these scenarios is a state with zero density, an infinite scale factor and an infinite $\dot{H}$, again of opposite signs. These scenarios are similar to the Big Shock, but in the phantom variant of the matter density.
The $X$ scenario realizes the Big Bang scenario, but with a final stage of de Sitter expansion.

As in the case of model \rf{HEGB}, all $I-X$ scenarios for model \rf{H37} corresponding to $\veps=-1$ can be analyzed in analogy with the scenarios for $\veps=+1$. But at the same time, they also have a number of paradoxical properties. Their analysis is of purely mathematical interest and is therefore not presented here.

\begin{figure}
    \centering
    \includegraphics[width=0.8\textwidth]{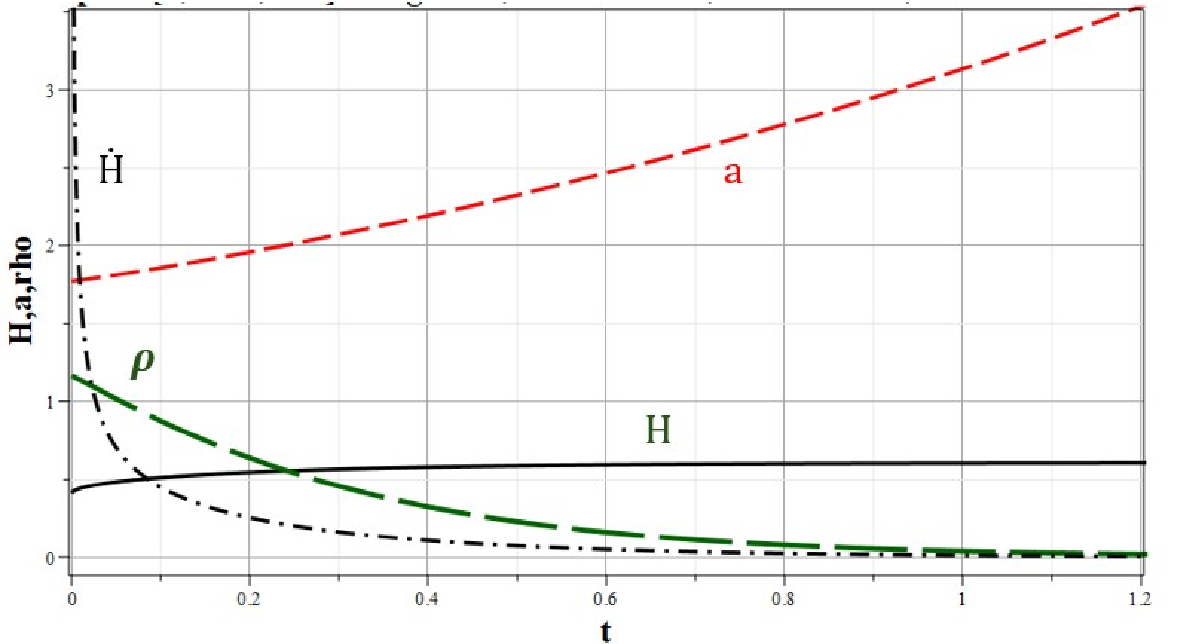}\\
    \textbf{a} \\ \vspace{0.2cm}
    \includegraphics[width=0.8\textwidth]{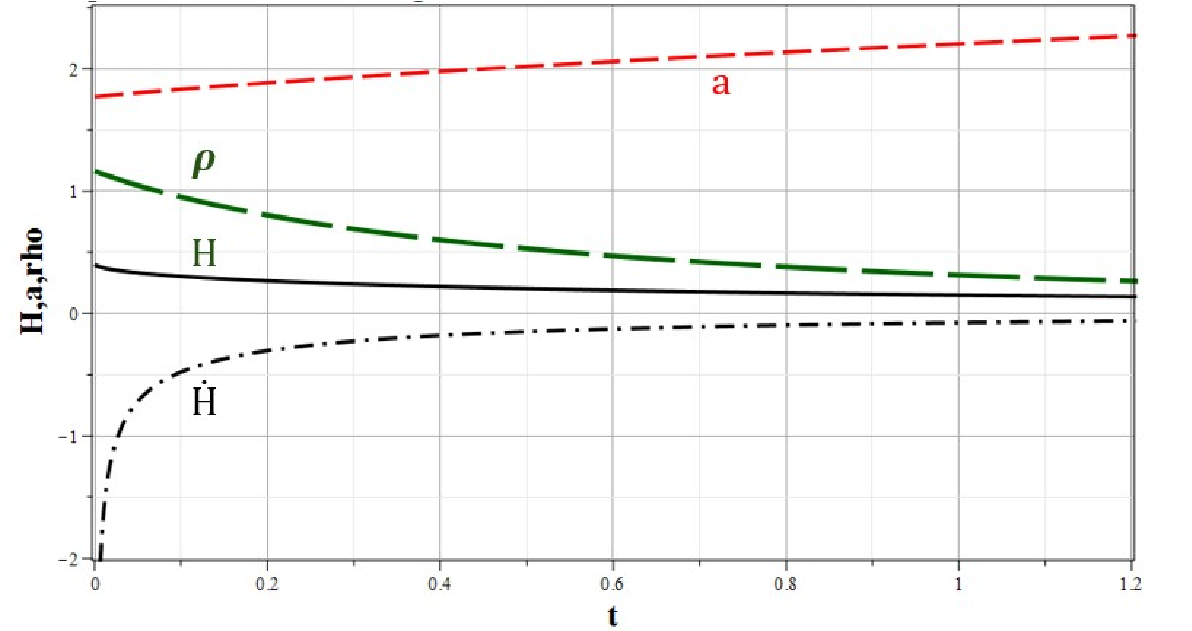}\\
    \textbf{b}
    \caption{Time evolution of $\dot{H},~H,~a,~\rho$ for the model \rf{H37} with the parameters of Fig.~\ref{Fig3} and initial conditions: a) $H_0=0.41$, b) $H_0=0.40$ }
    \label{Fig31}
    \end{figure}

By analogy with the models \eqref{HEGB}, Fig. \ref{Fig31} shows the time evolution of the main physical quantities of the model for the values of the parameters $\ga_i$ corresponding to the phase diagram in Fig. \ref{Fig3}a. The initial values $H(0)=H_0$ and $a(0)=a(H_0)$ are chosen so that in the case a) the evolution follows the Big Shock scenario $VII$, and in the case b) the scenario $VI$. A singular point of the model lies between $H_0=0.40$ and $H_0=0.41$. The plots illustrate the characteristic features of the evolution in the corresponding scenarios.

\section{Conclusion}

The classification of possible scenarios of cosmological evolution within the framework of Lovelock's theory of gravity with flat FRW space indicates a number of important features that are not found in models based on Einstein's general relativity. These features arise due to the additional elements of the Lagrangian that determine the geometry of spacetime and its evolution.

I. The most interesting and important features of Lovelock's cosmological models are that among the evolution scenarios in the case of the presence of singular points $H=\pm s_i$ of the system \rf{dotH}, there are "Big Shock"\ scenarios that replace the Big Bang scenarios. As shown in this work, the Big Bang scenarios are present in all Lovelock models in the case of positive values of all parameters $\alpha_i$. But in the case where some of the parameters $\alpha_i$ are negative and when in Lovelock models the density of the medium turns to zero for some values of the Hubble parameter, one of the possible scenarios of evolution turns out to be the Big Shock scenario. In this scenario, the Universe begins to evolve from a state with a finite value of the scale factor and a finite value of the density of matter, but with an infinite value of the rate of change of the Hubble parameter in time. These scenarios, from a physical point of view, look more rational than the Big Bang scenarios with respect to the initial data.
At the same time, such scenarios within the framework of Lovelock theory appear only in the dimensions of space-time $N>4$, which requires a revision of the general ideas about physical reality formulated in the general theory of relativity.

II. In addition to scenarios of the Big Shock and Big Bang type, among the possible scenarios in Lovelock models there are exotic scenarios. In the presence of the singular points $\pm s_i$ and of fixed points other than $H=0$, among the evolution scenarios there are scenarios with negative density of matter, which indicates a natural connection of Lovelock models with models of phantom scalar fields in the general theory of relativity. Therefore, it is important to further investigate the models discussed in this paper by incorporating matter in the form of scalar fields.

III. It should also be noted that among the discovered scenarios of the universe's evolution in the Lovelock theory, there are scenarios with an initial state of the universe having zero density and a zero scale factor, but only for $\veps=-1$. In higher orders of Lovelock theory, however, such situations are not excluded for $\veps=+1$ either. In these scenarios, the universe evolves into a state with a finite density and a finite scale factor. These scenarios also appear when $N>4$. From the point of view of physics, the existence of such scenarios seems to be acceptable, if we assume that the matter in the universe in these scenarios is generated directly from the geometry without any real material basis. The very possibility of such scenarios requires further research.

Thus, it can be stated that within the framework of Lovelock's cosmological models with flat space, a new set of scenarios for the evolution of the universe is discovered, which is not present in Einstein's general relativity. This allows us to consider Lovelock's theory of gravity as a justified development of the cosmological theory, which has recently faced difficulties in explaining a number of observed phenomena, as discussed in the introduction.

    \section*{Appendix A: Integration of equation \rf{dotadZ}}
    For the analytical integration of equation \rf{dotadZ} we use the partial-fraction decomposition
    \beq{DefiP2row}
        \frac{1}{ZP_2(Z)} = \frac{1}{A_2}\left(\frac{K_0}{Z} + \sum\limits_{i=1}^{n-1}\frac{K_i}{Z-Z_i}\right),
    \eeq
    where $Z_i=h^2_i$ are the roots of $P_2$, $A_2$ is its leading coefficient, $P_2(Z)=A_2\prod_{i=1}^{n-1}(Z-Z_i)$, and the constants $K_0,K_1,\ldots,K_{n-1}$ are found by reducing the right-hand side to a common denominator and equating coefficients of the resulting polynomial identity, valid for all $Z$:
    \beq{EqKi}
         \sum\limits_{i=0}^{n-1}K_i\prod\limits_{j=0,\,j\not=i}^{n-1}(Z-Z_j)=1,
    \eeq
    with the notation $Z_0=0$. In particular, for $n=2$ the identity yields the system
    $$
        K_0+K_1=0,~~K_0Z_1=-1,
    $$
    so that $K_0=-1/Z_1,~~K_1=1/Z_1$. For $n=3$ the identity \rf{EqKi} takes the form
    $$
         K_0(Z-Z_1)(Z-Z_2) + K_1(Z-Z_2)Z+K_2(Z-Z_1)Z=1.
    $$
    Equating the coefficients of the powers of $Z$, we find
    $$
         K_0+K_1+K_2=0,~~K_0(Z_1+Z_2)+K_1Z_2+K_2Z_1=0,~~K_0Z_1Z_2 = 1,
    $$
    whence
    \beq{DefR12}
        K_0=\frac{1}{Z_1Z_2},~~K_1=\frac{1}{Z_1(Z_1-Z_2)},~~K_2=-\frac{1}{Z_2(Z_1-Z_2)}.
    \eeq

    Substituting the decomposition \eqref{DefiP2row} into \rf{dotadZ} and writing $P_1(Z)=\sum_{j=0}^{n-1}\gb_jZ^j$ with $\gb_0=-(N-2)$, $\gb_j=\alpha_{j+1}\,{}^{(j+1)}k_{12}$, we obtain
    \beq{Intlna}
     \ln(a/a_0) = -\frac{M}{A_2}\, I(Z), ~~M=\frac{1}{2(1+\go)},
    \eeq
    where
    \beq{DefI}
       I(Z)= \int\left(\frac{K_0}{Z} + \sum\limits_{i=1}^{n-1}\frac{K_i}{Z-Z_i}\right)\sum\limits_{j=0}^{n-1}\gb_{j}Z^j\, dZ.
    \eeq
    To compute $I(Z)$ we use the identity
    \beq{DefIj}
        \frac{Z^j}{Z-Z_i}=
        \frac{Z^j-Z_i^j}{Z-Z_i}+\frac{Z_i^j}{Z-Z_i}=R_j(Z,Z_i)+\frac{Z_i^j}{Z-Z_i},
    \eeq
    where
    $$
         R_j(Z,Z_i) = \sum\limits_{k=0}^{j-1}Z^{k}Z_i^{j-1-k},
    $$
    in particular
    $$
         R_1 = 1,~~R_2=Z+Z_i,~~R_3=Z^2+ZZ_i+Z_i^2,\ldots
    $$
    Integrating term by term, we get
    $$
         \int\frac{Z^j}{Z-Z_i}\,dZ = T_j(Z,Z_i) + Z_i^j\ln|Z-Z_i|,
    $$
    where
    $$
         T_j(Z,Z_i) = \sum\limits_{k=1}^{j}\frac{Z^{k}}{k}Z_i^{j-k}.
    $$
    As a result, we obtain
    \beqa{EqIlna} \nonumber
	   && I(H^2) = K_0\Big(\gb_0\ln H^2
       +\sum\limits_{j=1}^{n-1}\gb_j T_j(H^2,0)\Big)+ \\
       && +\sum\limits_{i=1}^{n-1}K_i\left(P_1(Z_i)\ln|H^2-Z_i| + \sum\limits_{j=1}^{n-1}\gb_j T_j(H^2,Z_i)\right),
   \eeqa
   where we used $P_1(Z_i) = \sum_{j=0}^{n-1}\gb_j Z_i^j$ to collect the logarithmic terms. Together with \eqref{Intlna} this reproduces \eqref{Sola} of the main text.

\section*{Appendix B: A differential identity for $P_1$ and $P_2$, root interlacing and phantom-free intervals}

The polynomials $P_1(Z)$ and $P_2(Z)$, $Z=H^2$, are not independent: they are connected by the differential identity
\beq{BianchiPP}
     (N-1)\,P_1(Z) = 2\,\frac{d}{dZ}\Big[Z\,P_2(Z)\Big].
\eeq
The identity follows directly from the continuity equation. Indeed, the density is an explicit function of $Z$, $\rho = -ZP_2(Z)$ \eqref{DefrhoH}, so that $\dot\rho = \rho'(Z)\,2H\dot H$, while the field equation \eqref{flatF1} gives $\rho + p = P_1(H^2)\dot H$. Substituting both expressions into the continuity equation $\dot\rho + (N-1)H(\rho+p)=0$ and canceling the common factor $2H\dot H$, we obtain $\rho'(Z) = -\tfrac{N-1}{2}P_1(Z)$, which is \eqref{BianchiPP}. At the level of the combinatorial coefficients, \eqref{BianchiPP} is equivalent to the per-order relation
\beq{Bianchik}
     {}^{(i)}k_{12} = \frac{2i}{N-1}\,{}^{(i)}k_{21},\qquad i=2,\ldots,n,
\eeq
which can also be verified directly from the definitions \eqref{eqK}: after dividing by common factorials, it reduces to the elementary identity $2(N-i-1) + (N-1-2i)(2i-2) = 2i(N-2i)$.

The identity \eqref{BianchiPP} has several immediate consequences.

\textbf{1. Interlacing of the roots.} If $Z_*>0$ is a common root of $P_1$ and $P_2$, then \eqref{BianchiPP} yields $0=(N-1)P_1(Z_*) = 2Z_*P_2'(Z_*)$, hence $P_2'(Z_*)=0$ and $Z_*$ is a \emph{multiple} root of $P_2$. Therefore, whenever the positive roots of $P_2$ are simple --- which holds for generic values of the couplings $\alpha_i$ --- the sets of singular points $\{s_i\}$ and of fixed points $\{h_i\}$ are disjoint. Coincidences occur only on the discriminant locus of $P_2$, where a singular barrier merges with a fixed point; for $N=7$, $n=3$ this locus is the curve $\alpha_3 = 3\alpha_2^2/2$ discussed in Sec.~\ref{SecCubic} (at this point the double root of $P_2$ coincides with a root of $P_1$, in full agreement with the general statement above).

\textbf{2. The first barrier precedes the first fixed point.} Let $h_1^2$ be the smallest positive root of $P_2$. Since $P_2(0) = -(N-1)(N-2)/2 < 0$, the polynomial $P_2$ is non-positive on $[0,h_1^2]$ and $P_2'(h_1^2)\ge 0$. Then \eqref{BianchiPP} gives $(N-1)P_1(h_1^2) = 2h_1^2P_2'(h_1^2)\ge 0$, and since $P_1(0)=-(N-2)<0$, the polynomial $P_1$ has a root $s_1^2\in(0,h_1^2]$, with equality possible only in the degenerate case $P_2'(h_1^2)=0$. Hence for generic couplings
$$
     s_1 < h_1 ,
$$
and the entire central basin $(-s_1,s_1)$ lies in the region $\rho(H) = -H^2P_2(H^2)>0$: \emph{the central basin is free of phantom intervals for arbitrary admissible values of the couplings $\alpha_i$}.

\textbf{3. Pure Lovelock order.} For the pure theory of order $n$, $P_1 = -(N-2)+\alpha_n{}^{(n)}k_{12}Z^{n-1}$ and $P_2 = -(N-1)(N-2)/2+\alpha_n{}^{(n)}k_{21}Z^{n-1}$, positive roots exist only for $\alpha_n<0$ (with $N\ge 2n+1$), and \eqref{Bianchik} fixes their ratio universally:
$$
     \frac{h_1^{2}}{s_1^{2}} = n^{1/(n-1)} ,
$$
independently of $\alpha_n$ and $N$. In particular, $h_1^2 = 2s_1^2$ for the Einstein--Gauss--Bonnet theory, in agreement with the explicit values of Sec.~\ref{SecEGB}.

\textbf{4. A sufficient condition for the absence of phantom intervals.} All the coefficients ${}^{(i)}k_{21}$ are negative for $N\ge 2i+1$. Consequently, if all $\alpha_i\ge 0$, then $P_{20}(Z)\le 0$ for $Z\ge0$ and $P_2(Z)\le -(N-1)(N-2)/2<0$ everywhere, so that $\rho(H)>0$ for all $H\neq 0$ and phantom intervals are absent altogether. Both the phantom intervals and the singular barriers (positive roots of $P_1$) thus require at least one negative Lovelock coupling.

\bigskip
\noindent\textit{Dedicated to the memory of Professor Naresh Dadhich. The approach to the Lovelock field equations based on the independent components of the Riemann tensor, which underlies the present analysis and the preceding works of one of the authors (A.V.N.), grew out of his suggestion made during his visit to the University of KwaZulu-Natal.}

   \begin{acknowledgments}
   The authors are grateful to the anonymous referee for the careful reading of the manuscript and for the valuable comments, which allowed us to improve the paper.
   The work of V.M.Z. and S.V.C. was performed within the framework of Supplementary Agreement No. 073-03-2026-035/1 dated 02/21/2026 to the Agreement on the Provision of Subsidies to a Federal Budgetary or Autonomous Institution for the Financial Support of the State Assignment for the Provision of Public services (works) No. 073-03-2026-035 dated 01/23/2026, concluded between the Federal State Budgetary Educational Institution of Higher Education "Ulyanovsk State Pedagogical University named after I. N. Ulyanov"\ and the Ministry of Education of the Russian Federation.
   
\end{acknowledgments}

\bigskip
\noindent\textbf{Data Availability Statement.} This is a purely theoretical study; no new data were created or analysed, and data sharing is not applicable to this article.

\noindent\textbf{Conflict of interest.} The authors declare that they have no conflict of interest.


\begin{thebibliography}{99}

\bibitem{1} D.~Lovelock, J. Math. Phys. \textbf{12}, 498 (1971). doi:10.1063/1.1665613
\bibitem{2} N.~Dadhich, Pramana \textbf{74}, 875 (2010). doi:10.1007/s12043-010-0080-1
\bibitem{3} S.~Nojiri, S.~D. Odintsov, Int. J. Geom. Methods Mod. Phys. \textbf{4}, 115 (2007). doi:10.1142/S0219887807001928
\bibitem{NO2011} S.~Nojiri, S.~D. Odintsov, Phys. Rep. \textbf{505}, 59 (2011); arXiv:1011.0544. doi:10.1016/j.physrep.2011.04.001
\bibitem{NOO2017} S.~Nojiri, S.~D. Odintsov, V.~K. Oikonomou, Phys. Rep. \textbf{692}, 1 (2017); arXiv:1705.11098. doi:10.1016/j.physrep.2017.06.001
\bibitem{4} S.~Capozziello, M.~De Laurentis, Phys. Rep. \textbf{509}, 167 (2011). doi:10.1016/j.physrep.2011.09.003
\bibitem{5} R.~G. Cai, L.~M. Cao, Phys. Rev. D \textbf{79}, 024012 (2009). doi:10.1103/PhysRevD.79.024012
\bibitem{6} S.~Chakraborty, N.~Dadhich, Phys. Dark Universe \textbf{30}, 100658 (2020). doi:10.1016/j.dark.2020.100658
\bibitem{7} A.~V. Nikolaev, Eur. Phys. J. C \textbf{85}, 25 (2025). doi:10.1140/epjc/s10052-024-13695-5
\bibitem{8} S.~D. Maharaj, N.~Naidoo, G.~Amery, K.~S. Govinder, Eur. Phys. J. C \textbf{83}, 333 (2023). doi:10.1140/epjc/s10052-023-11513-y
\bibitem{9} S.~Naicker, S.~D. Maharaj, B.~P. Brassel, Gen. Rel. Gravit. \textbf{55}, 116 (2023). doi:10.1007/s10714-023-03157-w
\bibitem{10} S.~Naicker, S.~D. Maharaj, B.~P. Brassel, Eur. Phys. J. C \textbf{83}, 343 (2023). doi:10.1140/epjc/s10052-023-11483-1
\bibitem{11} A.~V. Nikolaev, S.~D. Maharaj, Eur. Phys. J. C \textbf{80}, 7 (2020).
\bibitem{Zh01JETP} V.~M. Zhuravlev, J. Exp. Theor. Phys. \textbf{93}, 903 (2001). doi:10.1134/1.1427102
\bibitem{Vikman2005} A.~Vikman, Phys. Rev. D \textbf{71}, 023515 (2005); arXiv:astro-ph/0407107. doi:10.1103/PhysRevD.71.023515
\bibitem{CaiQuintom2010} Y.-F. Cai, E.~N. Saridakis, M.~R. Setare, J.-Q. Xia, Phys. Rep. \textbf{493}, 1 (2010); arXiv:0909.2776. doi:10.1016/j.physrep.2010.04.001
\bibitem{BL89} N.~N. Bautin, E.~A. Leontovich, \textit{Methods and Techniques of Qualitative Analysis of Dynamical Systems in the Plane} (Nauka, Moscow, 1989).
\bibitem{Bogoyavlensky1980} O.~I. Bogoyavlensky, \textit{Methods of the Qualitative Theory of Dynamical Systems in Astrophysics and Gas Dynamics} (Nauka, Moscow, 1980).
\bibitem{Belinsky1985} V.~A. Belinsky, L.~P. Grishchuk, I.~M. Khalatnikov, Ya.~B. Zeldovich, Phys. Lett. B \textbf{155}, 232 (1985). doi:10.1016/0370-2693(85)90644-6
\bibitem{Copeland2006} E.~J. Copeland, M.~Sami, S.~Tsujikawa, Int. J. Mod. Phys. D \textbf{15}, 1753 (2006). doi:10.1142/S021827180600942X
\bibitem{BCNO2012} K.~Bamba, S.~Capozziello, S.~Nojiri, S.~D. Odintsov, Astrophys. Space Sci. \textbf{342}, 155 (2012); arXiv:1205.3421. doi:10.1007/s10509-012-1181-8
\bibitem{Ign1} Yu.~G. Ignat'ev, \textit{Classical Cosmology and Dark Energy} (Kazan University Press, Kazan, 2016).
\bibitem{Ign2} Yu.~G. Ignat'ev, Space Time Fundam. Interact., no.~3, 17 (2016).
\bibitem{Zh16STFI} V.~M. Zhuravlev, Space Time Fundam. Interact., no.~4, 39 (2016).
\bibitem{ZhPP11GC} V.~M. Zhuravlev, T.~V. Podymova, E.~A. Pereskokov, Gravit. Cosmol. \textbf{17}, 101 (2011). doi:10.1134/S0202289311020204
\bibitem{iug-2} Yu.~G. Ignat'ev, A.~R. Samigullina, Gravit. Cosmol. \textbf{31}, 1 (2025); arXiv:2410.10703. doi:10.1134/S0202289324700439
\bibitem{iug-3} Yu.~G. Ignat'ev, I.~A. Kokh, Gravit. Cosmol. \textbf{30}, 426 (2024). doi:10.1134/S0202289324700324
\bibitem{iug-4} Yu.~G. Ignat'ev, D.~Yu. Ignatyev, Gravit. Cosmol. \textbf{26}, 29 (2020); arXiv:2005.14010.
\bibitem{iug-5} Yu.~G. Ignat'ev, Theor. Math. Phys. \textbf{219}, 688 (2024); arXiv:2307.13761. doi:10.1134/S0040577924040123
\bibitem{zhch-2020} V.~Zhuravlev, S.~Chervon, Universe \textbf{6}, 195 (2020). doi:10.3390/universe6110195
\bibitem{Starobinsky} A.~A. Starobinsky, Phys. Lett. B \textbf{91}, 99 (1980). doi:10.1016/0370-2693(80)90670-X
\bibitem{Dehghani2009} M.~H. Dehghani, N.~Farhangkhah, Phys. Lett. B \textbf{674}, 243 (2009). doi:10.1016/j.physletb.2009.03.045
\bibitem{Deser2012} S.~Deser, J.~Franklin, Class. Quantum Grav. \textbf{29}, 072001 (2012). doi:10.1088/0264-9381/29/7/072001
\bibitem{Dadhich2013} N.~Dadhich, J.~M. Pons, K.~Prabhu, Gen. Rel. Gravit. \textbf{45}, 1131 (2013). doi:10.1007/s10714-013-1514-0
\bibitem{Padmanabhan2013} T.~Padmanabhan, D.~Kothawala, Phys. Rep. \textbf{531}, 115 (2013). doi:10.1016/j.physrep.2013.05.007
\bibitem{Chakraborty2015} S.~Chakraborty, J. High Energy Phys. \textbf{2015}, 029 (2015). doi:10.1007/JHEP08(2015)029
\bibitem{Bueno2016} P.~Bueno, P.~A. Cano, O.~Lasso A., P.~F. Ram\'\i rez, J. High Energy Phys. \textbf{2016}, 028 (2016). doi:10.1007/JHEP04(2016)028
\bibitem{Concha2017} P.~Concha, E.~Rodr\'\i guez, Phys. Lett. B \textbf{774}, 616 (2017). doi:10.1016/j.physletb.2017.10.019
\bibitem{Pavluchenko2024} S.~A. Pavluchenko, Universe \textbf{10}, 429 (2024). doi:10.3390/universe10110429
\bibitem{Bousder2023} M.~Bousder, A.~Riadsolh, M.~El Belkacemi, H.~Ez-Zahraouy, Ann. Phys. \textbf{458}, 169441 (2023). doi:10.1016/j.aop.2023.169441
\bibitem{BrasselSubmitted} B.~P. Brassel, \textit{Vacuum dark energy and equation of state for a Lovelock--FLRW universe with effective thermodynamics}, submitted (2025).
\bibitem{SinghBrasselMaharaj2025} S.~Singh, B.~P. Brassel, S.~D. Maharaj, Universe \textbf{11}, 155 (2025). doi:10.3390/universe11050155
\bibitem{BrasselSinghMaharaj2025} B.~P. Brassel, S.~Singh, S.~D. Maharaj, Ann. Phys. \textbf{482}, 170234 (2025). doi:10.1016/j.aop.2025.170234

\bibitem{Arroyo2026}
A.~Arroyo, R.~Cordero, G.~Cruz, E.~Rojas, Class. Quantum Grav. \textbf{43}, 055008 (2026); arXiv:2509.05920.

\bibitem{MyrzakulovKoussourGogoi2023}
N.~Myrzakulov, M.~Koussour, D.J.~Gogoi, Phys. Dark Univ. \textbf{42}, 101268 (2023); arXiv:2306.13218.

\bibitem{Kong2025}
Shi-Bei Kong, Fortschr. Phys. \textbf{73}, e70052 (2025).

\bibitem{Esmakhanova2011}
K.~Esmakhanova, N.~Myrzakulov, G.~Nugmanova, Y.~Myrzakulov, L.~Chechin, R.~Myrzakulov, Int. J. Mod. Phys. D \textbf{20}, 2419 (2011); arXiv:1104.3705.

\bibitem{NOT2005}
S.~Nojiri, S.D.~Odintsov, S.~Tsujikawa, Phys. Rev. D \textbf{71}, 063004 (2005); arXiv:hep-th/0501025.

\bibitem{Barrow2004}
J.D.~Barrow, Class. Quantum Grav. \textbf{21}, L79 (2004); arXiv:gr-qc/0403084.

\bibitem{deHaro2023}
J.~de Haro, S.~Nojiri, S.~D. Odintsov, V.~K. Oikonomou, S.~Pan, Phys. Rep. \textbf{1034}, 1 (2023); arXiv:2309.07465. doi:10.1016/j.physrep.2023.09.003

\bibitem{FJL2004}
L.~Fern\'andez-Jambrina, R.~Lazkoz, Phys. Rev. D \textbf{70}, 121503(R) (2004); arXiv:gr-qc/0410124.

\bibitem{KMPT2010}
I.V.~Kirnos, A.N.~Makarenko, S.A.~Pavluchenko, A.V.~Toporensky, Gen. Relativ. Gravit. \textbf{42}, 2633 (2010); arXiv:0906.0140.

\bibitem{NaickerMaharajBrassel2026}
S.~Naicker, S.D.~Maharaj, B.P.~Brassel, Gen. Relativ. Gravit. \textbf{58}, 24 (2026).

\bibitem{MaharajNaickerBrassel2026}
S.D.~Maharaj, S.~Naicker, B.P.~Brassel, Class. Quantum Grav. \textbf{43}, 015010 (2026).

\bibitem{NaickerBrasselMaharaj2026}
S.~Naicker, B.P.~Brassel, S.D.~Maharaj, Ann. Phys. \textbf{493}, 170589 (2026).

\bibitem{MaharajBrasselSingh2026}
S.D.~Maharaj, B.P.~Brassel, S.~Singh, K.S.~Govinder, Nucl. Phys. B \textbf{1023}, 117307 (2026).

\bibitem{MaharajGovinder2025}
S.D.~Maharaj, K.S.~Govinder, Gen. Relativ. Gravit. \textbf{57}, 11 (2025).

\bibitem{BogadiEtAl2026}
R.S.~Bogadi, G.~Leon, M.~Govender, K.S.~Govinder, S.~Maharaj, A.~Paliathanasis, Gen. Relativ. Gravit. \textbf{58}, 40 (2026).

\bibitem{Zwiebach1985}
B.~Zwiebach, Phys. Lett. B \textbf{156}, 315 (1985).

\bibitem{Zumino1986}
B.~Zumino, Phys. Rep. \textbf{137}, 109 (1986).

\bibitem{BoulwareDeser1985}
D.G.~Boulware, S.~Deser, Phys. Rev. Lett. \textbf{55}, 2656 (1985).

\bibitem{AokiMotohashi2020}
K.~Aoki, H.~Motohashi, JCAP \textbf{08} (2020) 026; arXiv:2001.06756.

\bibitem{KitauraWheeler1991}
T.~Kitaura, J.T.~Wheeler, Nucl. Phys. B \textbf{355}, 250 (1991).
\bibitem{Duchaniya2024} L.~K. Duchaniya, B.~Mishra, I.~V. Fomin, S.~V. Chervon, Class. Quantum Grav. \textbf{41}, 235016 (2024). doi:10.1088/1361-6382/ad8a13
\bibitem{CLM2014} S.~Capozziello, F.~S.~N. Lobo, J.~P. Mimoso, Phys. Lett. B \textbf{730}, 280 (2014); arXiv:1312.0784. doi:10.1016/j.physletb.2014.01.066
\bibitem{C02} R.~R. Caldwell, Phys. Lett. B \textbf{545}, 23 (2002); arXiv:astro-ph/9908168. doi:10.1016/S0370-2693(02)02589-3
\bibitem{GPRS06} R.~Gannouji, D.~Polarski, A.~Ranquet, A.~A. Starobinsky, JCAP \textbf{0609}, 016 (2006). doi:10.1088/1475-7516/2006/09/016; arXiv:astro-ph/0606287.
\end{thebibliography}
\end{document}